\documentclass[12pt]{article}
\usepackage[a4paper,margin=1in]{geometry}
\usepackage[titletoc,title]{appendix}
\usepackage{changepage}
\usepackage[utf8x]{inputenc}
\usepackage{textcomp,marvosym}
\usepackage{fixltx2e}
\usepackage{amsmath,amssymb}
\usepackage{cite}
\usepackage{nameref,hyperref}
\usepackage{authblk}
\usepackage{algorithm}
\usepackage{algpseudocode}
\usepackage{dsfont}
\usepackage{xcolor}
\usepackage{enumitem}
\usepackage{caption,setspace}
\usepackage{subcaption}
\usepackage{pdflscape}
\usepackage{graphicx}
\usepackage{placeins}
\usepackage{mathtools}
\usepackage{authblk}
\usepackage{epstopdf,float}
\DeclareMathOperator\arctanh{arctanh}
\usepackage{color}
\def\cbl{\color{black}}
\def\cb{\color{black}}

\begin{document}
\title{Non-vanishing sharp-fronted travelling wave solutions of the Fisher-Kolmogorov model}

\author[1]{Maud El-Hachem}
\author[1]{Scott W McCue}
\author[1*]{Matthew J Simpson}
\affil[1]{School of Mathematical Sciences, Queensland University of Technology, Brisbane, Australia.}
\affil[*]{Corresponding author: Matthew J Simpson, matthew.simpson@qut.edu.au}

\maketitle

%%%% Abstract text to be placed here %%%%%%%%%%%%
\begin{abstract}
The Fisher-KPP model, and generalisations thereof, are simple reaction-diffusion models of biological invasion that assume individuals in the population undergo linear diffusion with diffusivity $D$, and logistic proliferation with rate $\lambda$. For the Fisher-KPP model, biologically-relevant initial conditions lead to long-time travelling wave solutions that move with speed $c=2\sqrt{\lambda D}$.  Despite these attractive features, there are several biological limitations of travelling wave solutions of the Fisher-KPP model.  First, these travelling wave solutions do not predict a well-defined invasion front.  Second, biologically-relevant initial conditions lead to travelling waves that move with speed $c=2\sqrt{\lambda D} > 0$.  This means that, for biologically-relevant initial data, the Fisher-KPP model can not be used to study invasion with $c \ne 2\sqrt{\lambda D}$, or retreating travelling waves with $c < 0$.  Here, we reformulate the Fisher-KPP model as a moving boundary problem on $x < s(t)$ and show that this reformulated model alleviates the key limitations of the Fisher-KPP model.  Travelling wave solutions of the moving boundary problem predict a well-defined front, and can propagate with \textit{any} wave speed, $-\infty < c < \infty$.  Here, we establish these results using a combination of high-accuracy numerical simulations of the time-dependent partial differential equation, phase plane analysis and perturbation methods. All software required to replicate this work is available on  \href{https://github.com/ProfMJSimpson/TravellingWaves_ElHachem2021}{GitHub}.
\end{abstract}


\newpage

\section{Introduction}
\label{sec:intro}
The Fisher-Kolmogorov model, also known as the Fisher-KPP model, is a widely-used one-dimensional reaction-diffusion model that describes the spatial and temporal evolution of a population of motile and proliferative individuals with density $u(x,t)$~\cite{Fisher1937,Kolmogorov1937}.  Individuals in the population are assumed to undergo  diffusion with diffusivity $D$ and logistic proliferation with proliferation rate $\lambda$, and have a carrying capacity density $K$.

The Fisher-KPP model, and various extensions, have been used to study a range of biological phenomena including various applications in cell biology~\cite{Sherratt90,Swanson2003,Painter2003,Gatenby1996,Landman1998,Maini2004a,Maini2004b,Jin2016,Bitsouni2018,Warne2019} and ecology~\cite{Shigesada1995,Skellam1951,Steel1998,Kot03}.  From a mathematical point of view, the Fisher-KPP model is of high interest because it supports travelling wave solutions that have been widely studied using a range of mathematical techniques~\cite{Canosa1973,ElHachem2019,Murray02,Aronson1978}.  Despite the immense interest in travelling wave solutions of the Fisher-KPP model, there are various features of these solutions that are biologically unsatisfactory.  For example, travelling wave solutions of the Fisher-KPP model are smooth and without compact support, and $u(x,t) \to 0$ \cbl as $x \to \infty$\cb.  This means that these travelling wave solutions do not provide a clear way to model the motion of a well-defined invasion front~\cite{Maini2004a,Maini2004b}.  Furthermore, travelling wave solutions of the Fisher-KPP model that evolve from initial conditions with compact support lead to long-time travelling waves that move with speed $c = 2 \sqrt{\lambda D}$~\cite{Canosa1973,Murray02}.  Despite the fact that constant speed travelling wave-type behaviour can be observed and measured experimentally~\cite{Maini2004a,Maini2004b}, simply observing travelling wave-type behaviour does not verify the relationship $c = 2 \sqrt{\lambda D}$.  Another limitation of the Fisher-KPP model is that travelling wave solutions always lead to invading fronts with $c > 0$ and \cbl $\partial u(x,t) / \partial t > 0$. \cb In contrast, various applications in biology and ecology involve retreating fronts with $c < 0$ and \cbl $\partial u(x,t) / \partial t < 0$\cb~\cite{ElHachem2021a}, and these processes cannot be modelled using the Fisher-KPP model.

Various mathematical extensions have been proposed to overcome the biologically unsatisfactory features of the Fisher-KPP model. Perhaps the most widely known extension is to generalise the linear diffusion term in the Fisher-KPP model to a degenerate nonlinear diffusion term, giving rise to a model that is often called the Porous-Fisher model~\cite{Murray02,Sengers2007,Sanchez1994,Sanchez1995,Witelski1994,Witelski1995,McCue2019}.  The Porous-Fisher model leads to sharp-fronted travelling wave solutions that can be used to model the motion of a well-defined front, such as those that are often observed experimentally~\cite{Maini2004a,Maini2004b}.  With a nonlinear degenerate diffusivity $\mathcal{D}(u) = D u$, time-dependent solutions of the Porous-Fisher model with initial conditions that have compact support leads to travelling waves that move with speed $c = \sqrt{\lambda D / 2}$.  Again, experimental measurements of the wave speed \cbl do \cb not confirm the relationship $c = \sqrt{\lambda D / 2}$.  Similar to the Fisher-KPP model, the Porous-Fisher model cannot be used to study retreating fronts~\cite{ElHachem2021a}.   A second, less common approach to overcome the biologically unsatisfactory features of the Fisher-KPP model is to reformulate the model as a moving boundary problem on $x < s(t)$, where the density vanishes on the moving front, $u(s(t),t)=0$, meaning that this moving boundary problem gives rise to a well-defined front that is consistent with experimental observations.  This model, where the motion of $s(t)$ is given by a classical one-phase Stefan condition $\textrm{d} s(t) / \textrm{d} t = -\kappa \partial u(s(t),t) / \partial x$~\cite{Crank1987,Hill1987,Gupta2017}, has been called the Fisher-Stefan model~\cite{Du2010,Du2014a,Du2014b,ElHachem2019}.   While moving boundary problems of this type are most often used to study certain physical and industrial phenomena~\cite{Mitchell2014,BrosaPlanella2019,Dalwadi2020,BrosaPlanella2021}, they are also used to study biological processes, such as tumour spheroid growth and wound healing~\cite{Ward1997,Ward1999,Gaffney1999,Kimpton2013,Fadai2020,Jin2021}. Setting $\kappa > 0$ in the Fisher-Stefan model can lead to travelling wave solutions with $0 < c < 2\sqrt{\lambda D}$. Unlike either the Fisher-KPP or Porous-Fisher models, the Fisher-Stefan model can be used to model retreating travelling waves with $c < 0$ simply by setting $\kappa < 0$~\cite{ElHachem2021a}.  In summary, the Fisher-Stefan model can be used to study a wide range of travelling wave solutions with $-\infty < c < 2 \sqrt{\lambda D}$.  From this point of view, the Fisher-Stefan model is much more flexible than either the classical Fisher-KPP or Porous-Fisher models.

\begin{figure}[H]
	\centering
	\includegraphics[width=0.9\linewidth]{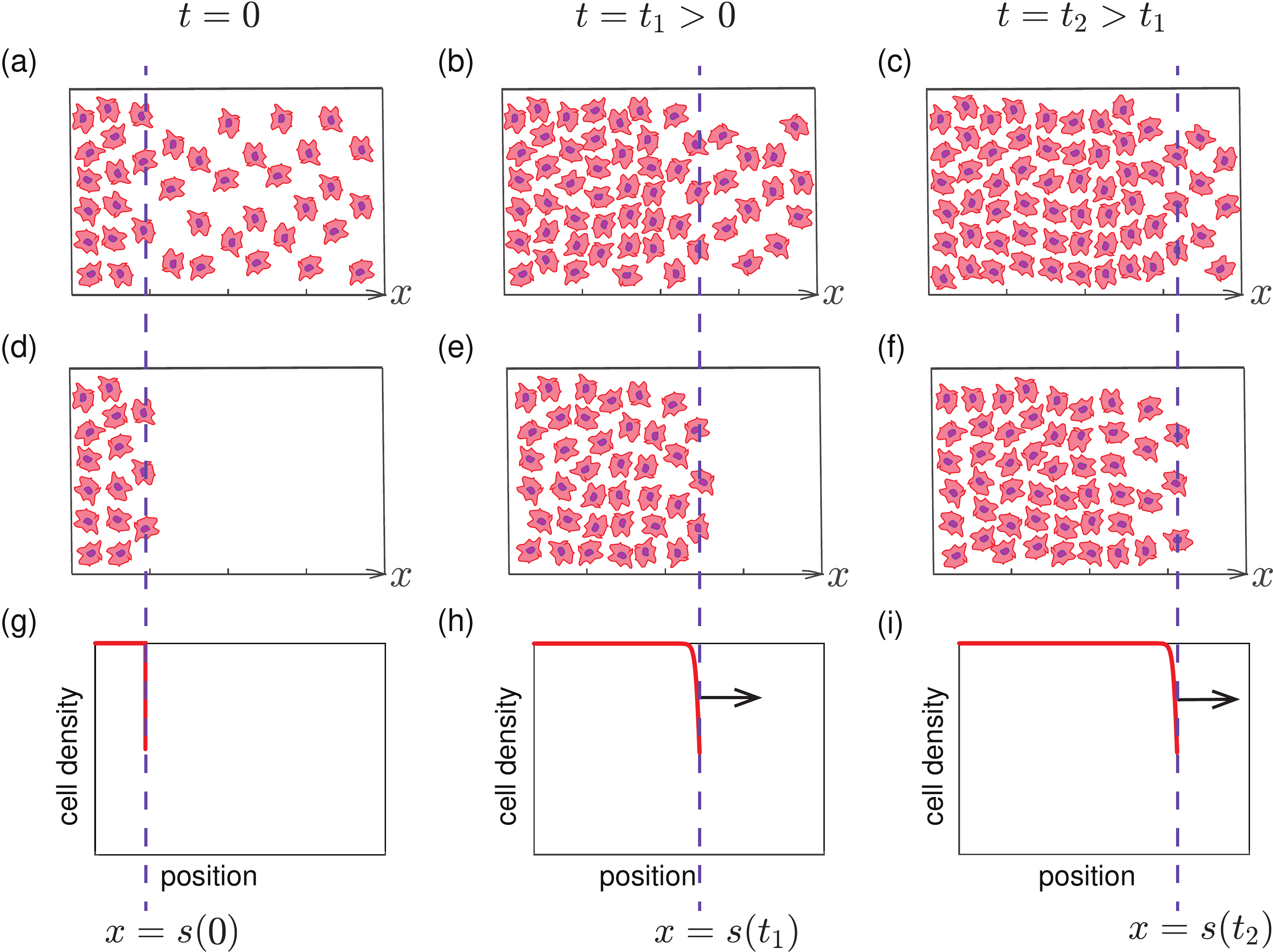}
	\caption{\textbf{Schematic showing two interpretations of the non-vanishing Stefan model of invasion.} (a)--(c) Evolution of a motile and proliferative cell population leading to an invading front moving into an initially occupied region. (d)--(f) Evolution of a motile and proliferative cell population leading to an invading front moving into an initially-vacant region.  (g)--(i) Both  schematics lead to an evolving density profile, moving in the positive $x$-direction with a non-vanishing,  sharp-front density profile. Each column, from left-to-right, shows snapshots at different values of time, $t=0, t_1$ and $t_2$, with $0 < t_1 < t_2$, and the position of the moving front, $x=s(t)$, is shown with three dashed vertical lines.}
	\label{fig:1}
\end{figure}

In this work we propose and analyse a generalisation of the Fisher-Stefan model that enables us to study travelling wave solutions with any wave speed,  $-\infty < c < \infty$.  This flexibility arises by generalising the boundary condition at the moving front, $x=s(t)$.  The usual Fisher-Stefan model involves setting $u(s(t),t)=0$ so that the solution vanishes at $x = s(t)$.  Here, we set $u(s(t),t)=u_{\textrm{f}}$, where $u_{\textrm{f}} \in [0,1)$ so that the density at the moving front is non-vanishing.  Of course, this generalisation simplifies to the usual Fisher-Stefan model when $u_{\textrm{f}}=0$. There are two different ways of motivating this kind of boundary condition, illustrated schematically in Figure \ref{fig:1} in the context of cellular invasion. \cbl First, in Figure \ref{fig:1}(a)--(c), we think of a population of motile and proliferative cells that give rise to an invading front moving into an existing background population of cells ahead of the moving boundary with  $u(s(t),t)=u_{\textrm{f}}$.  Recall that the Fisher-KPP model is often used to model the invasion of one population of cells, such as a tumour cell population, into a surrounding population of healthy cells by simply modelling the invading population~\cite{Sherratt90,Swanson2003,Maini2004a,Maini2004b,Jin2016,Bitsouni2018,Warne2019}  rather than explicitly modelling both populations~\cite{Gatenby1996,Landman1998,Painter2003,ElHachem2020,Elhachem2021}.  Our approach can be thought of as a hybrid approach where we deal only with a PDE for the invading population, but we explicitly model the impact of the surrounding tissue by choosing $u_{\textrm{f}} \in [0,1)$.  Second, in Figure \ref{fig:1}(d)--(f) we think of a population of motile and proliferative cells that give rise to an invading front that moves into empty space ahead of the moving boundary with $u(s(t),t)=u_{\textrm{f}}$ at the leading edge. \cb In both cases, the one-phase Stefan condition at $x=s(t)$ implies there is a local loss of the invading population at the leading edge.  Regardless of the motivation for this model, our interest is in modelling the behaviour of the invading population in the region $x < s(t)$.  While the schematic in Figure \ref{fig:1} is presented in terms of an invading front with $c > 0$, a similar schematic with very similar interpretations can be drawn for a retreating front with $c < 0$.

This work is organised as follows.  We first introduce time-dependent solutions of the partial differential equation (PDE) model where we demonstrate that late-time numerical solutions give rise to a range of invading and retreating travelling waves.  Following this numerical motivation, we show how these late-time PDE solutions are related to various trajectories in the classical Fisher-KPP phase plane~\cite{Murray02}.  Focusing on the phase plane, we then obtain a range of solutions describing various travelling wave phenomena, including exact solutions for stationary waves, $c=0$, and exact solutions for which the ordinary differential equation (ODE) governing the phase plane has the Painlev\'{e} property, $c = \pm 5/\sqrt{6}$~\cite{Ablowitz1979,Kaliappan1984,McCue2021}.  Building on these exact results, we then obtain various approximate perturbation solutions which allow us to study: (i) slowly invading or retreating travelling waves, $|c| \ll 1$; (ii) fast retreating travelling waves, $c \to -\infty$; and, (iii) fast invading travelling waves, $c \to \infty$.  At the outset, we acknowledge that one of the weaknesses of the Fisher-Stefan model is the lack of biological interpretation of the parameter $\kappa$ and a lack of methods for measuring this parameter. Our analysis overcomes this limitation since our exact and perturbation solutions allow us to relate $\kappa$ to the wave speed, $c$.  This is a useful outcome because experimental measurements of $c$ are relatively straightforward to obtain and so our analysis allows us to interpret such measurements of $c$ in terms of $\kappa$, \cbl given that the density $u_{\textrm{f}}$ of the population at the interface is also known, from experimental measurements.\cb

\section{Results and discussion}

\subsection{Mathematical model} \label{sec:mathmodel}
We begin by studying the numerical solutions of the following non-dimensional moving boundary problem~\cite{Du2010,Du2014a,Du2014b}
\begin{align}
&\dfrac{\partial u}{\partial t} =  \dfrac{\partial^2 u}{\partial x^2} +  u\left(1-u \right), \quad 0 < x<s(t), \label{eq:NondimPDE} \\
&\dfrac{\partial u(0,t)}{\partial x} = 0, \quad u(s(t),t) = u_\mathrm{f}, \quad \dfrac{\text{d}s(t)}{\text{d} t} = -\kappa \frac{\partial u(s(t),t)}{\partial x}, \label{eq:NondimBC}
\end{align}
where \cbl $u(x,t) \ge 0$ \cb is the population density~\cite{ElHachem2019}.  The length of the domain, $s(t)$, is determined as part of the solution through the classical one-phase Stefan condition. As we described in the Introduction, the key novelty here is to consider a non-vanishing boundary condition $u(s(t),t)= u_\mathrm{f} \in[0,1)$, which means that our model simplifies to the Fisher-Stefan model in the special case where  $u_\mathrm{f}=0$.  While our travelling wave analysis is valid on an infinite domain, we study time-dependent travelling waves by working with a sufficiently large finite domain, $0 < x < s(t)$. For all time-dependent PDE solutions we consider the initial condition ,
\begin{align}\label{eq:ICPDE}
u(x,0) =
\begin{cases}
& 1,  \qquad \qquad \qquad \qquad \qquad \qquad \quad 0 < x < \beta,\\
& \dfrac{(1- u_\mathrm{f})}{(s(0)-\beta)}(s(0)-x) +  u_\mathrm{f}, \quad  \beta < x < s(0),
\end{cases}
\end{align}
which is a ramp-shaped function for which we must specify values of $\beta>0$ and $s(0)$.  \cbl While all results in this work focus on this linear ramp function, there are many other options for $u(x,0)$.  The key property $u(x,0)$ is that we have $u(x,0) = 1$ near $x=0$, and $u(s(0),0)=u_{\textrm{f}}$ at the front.  Our choice of a linear ramp function is the simplest choice of initial condition to meet these properties, however other functional forms are possible, such as a nonlinear function of position.  Preliminary numerical experimentation (not shown) indicates that the long--time travelling wave solutions of the mathematical model do not on these details. \cb Note that when we study invading travelling waves we choose $s(0)=1$, whereas when we study retreating travelling waves we choose $s(0) \gg 1$~\cite{McCue2021b}.  Full details of the numerical method to solve this moving boundary problem are given in Appendix A, and MATLAB software to implement these algorithms are available on \href{https://github.com/ProfMJSimpson/TravellingWaves_ElHachem2021}{GitHub}.

\subsection{Time dependent PDE solutions}  \label{sec:timedependantsol}
\begin{figure}[H]
	\centering
	\includegraphics[height=0.9\textheight]{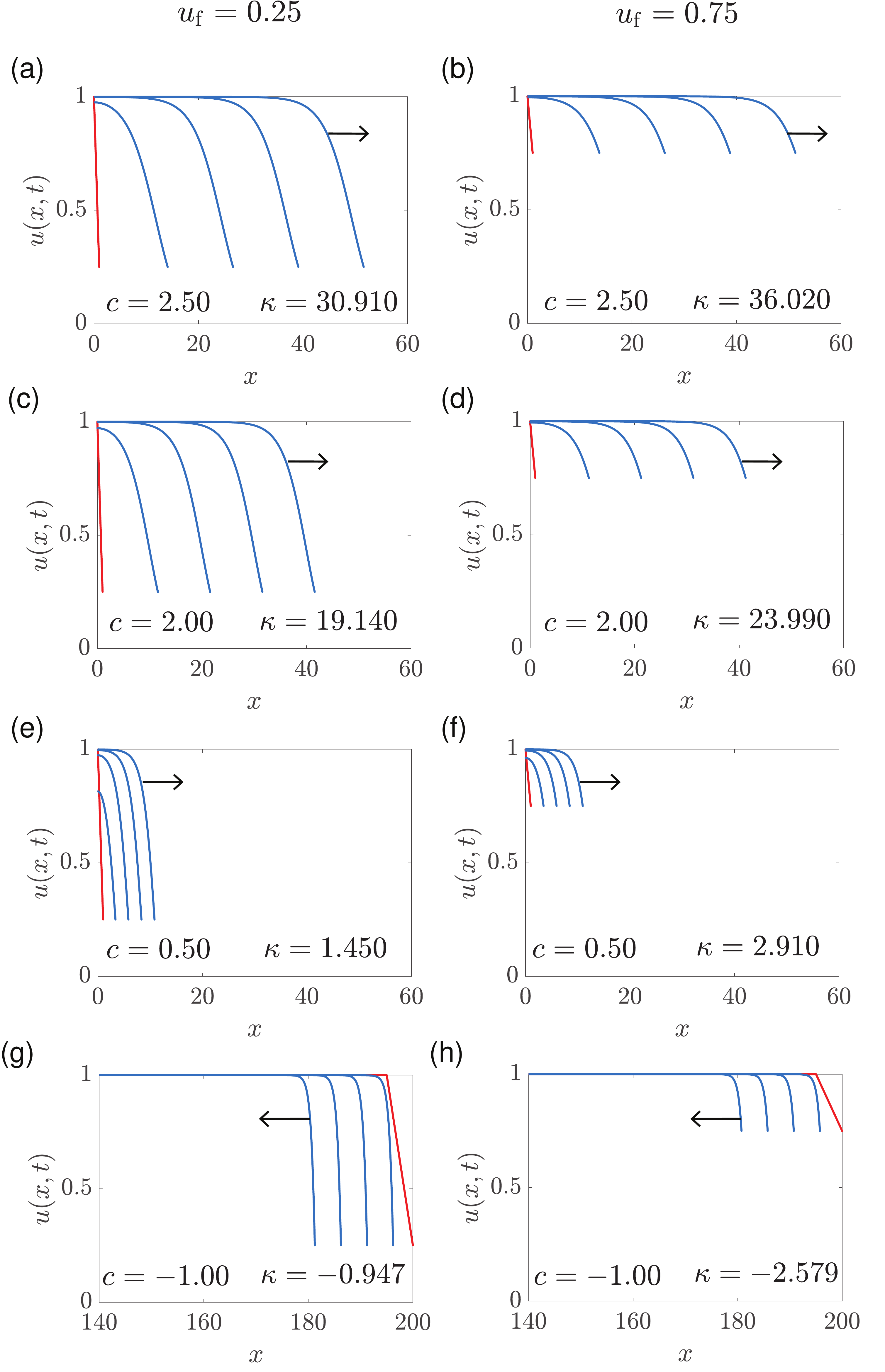}
	\caption{\textbf{Time-dependant solutions of Equations (\ref{eq:NondimPDE})--(\ref{eq:ICPDE}).} Density profiles $u(x,t)$ (blue) at times $t=5, 10, 15$ and $20$, evolving from the initial condition (red) with $s(0)=1$ and $\beta = 0$ in (a)--(f), and $s(0)=200$ and $\beta = 195$ in (g)--(h). Results in (a), (c) and (e) evolve into invading travelling wave solutions with $c=2.50,2.00$ and $0.50$, respectively.  Profiles in (a), (c) and (e) correspond to $u_\mathrm{f}=0.25$ while profiles in (b), (d) and (f) correspond to $u_\mathrm{f}=0.75$.  Results in (g) and (h) evolve into retreating travelling wave solutions, both with $c=-1.00$.  Profiles in (g) and (h) correspond to $u_\mathrm{f}=0.25$ and $u_\mathrm{f}=0.75$, respectively.  The values of $\kappa$ are given in each subfigure.}
	\label{fig:2}
\end{figure}

Numerical results in Figure \ref{fig:2} show the evolution of $u(x,t)$ for various choices of $\kappa$. In all cases we see that the initial condition rapidly evolves into a constant speed, constant shape travelling wave solution.  Results in the left column of Figure \ref{fig:2} involve $u_\mathrm{f} = 0.25$ while the results in the right column involve  $u_\mathrm{f}=0.75$, and we see in all cases that the density is non-vanishing at the front of the profile, $x=s(t)$.  Results in Figure \ref{fig:2}(a)--(f) involve setting $\kappa > 0$ meaning that the time-dependent PDE solutions evolve to invading travelling wave solutions with $c > 0$.  It is interesting to note that results in Figure \ref{fig:2}(e)--(f) involve travelling wave solutions with $c=0.50$, which is not possible with the usual nondimensional Fisher-KPP or Porous-Fisher models since travelling wave solutions for those models never move with such a slow wave speed~\cite{Murray02}.  Results in Figure \ref{fig:2}(g)--(h) involve $\kappa < 0$ and so lead to retreating travelling waves with $c < 0$.  Again, neither of these results are possible using the Fisher-KPP or Porous-Fisher models~\cite{ElHachem2021a}.  Now that we have provided numerical evidence of this range of late-time travelling wave behaviour in terms of the time-dependent PDE solutions, we will analyse these travelling wave solutions using the phase plane.

\subsection{Phase plane analysis} \label{sec:travwavesol}
In the usual way, we analyse travelling wave solutions by re-writing Equation (\ref{eq:NondimPDE}) in terms of the travelling wave coordinate, $z = x-ct$~\cite{Canosa1973,Murray02}.  We seek solutions of the form $u(x,t)=U(z)$ which leads to
\begin{equation} \label{eq:ODEUz}
\frac{\mathrm{d}^2 U}{\mathrm{d} z^2} + c\frac{\mathrm{d} U}{\mathrm{d} z} + U(1-U) = 0,  \quad  -\infty < z < 0,
\end{equation}
with boundary conditions
\begin{align}
U(-\infty) &= 1, \quad   U(0)= u_\mathrm{f}, \quad c = -\kappa \frac{\mathrm{d} U(0)}{\mathrm{d} z}, \label{eq:ODEBC}
\end{align}
where, for convenience, we have chosen $z=0$ to correspond to the moving boundary.

To proceed, we re-write Equation (\ref{eq:ODEUz}) as a first order dynamical system
\begin{align}
\frac{\text{d}U}{\text{d} z} & = V, \label{eq:ODEdU}\\
\frac{\text{d}V}{\text{d} z} &= -cV - U(1-U), \label{eq:ODEdV}
\end{align}
which defines the well-known phase plane associated with travelling wave solutions of the Fisher-KPP model~\cite{Canosa1973,Murray02}. Full details of how we obtain numerical trajectories in the phase plane are given in Appendix A. This phase plane involves two equilibrium points $(\bar{U},\bar{V}) = (0,0)$ and $(\bar{U},\bar{V})=(1,0)$.  Linearisation shows that $(\bar{U},\bar{V}) = (0,0)$ is a stable spiral if $c^2<4$, and a stable node if $c^2>4$, whereas $(\bar{U},\bar{V})= (1,0)$ is a saddle for all $c$.  Normally, in standard phase plane analysis of the Fisher-KPP model we reject travelling wave solutions with $c^2<4$ on physical grounds since the local behaviour about the origin implies that the density goes negative as the \cbl heteroclinic \cb trajectory between $(1,0)$ and $(0,0)$ spirals into the origin.  Here, we find that no such restriction is necessary as we will now explain.

\begin{landscape}
	\begin{figure}[H]
		\centering
		\includegraphics[height=0.70\textheight]{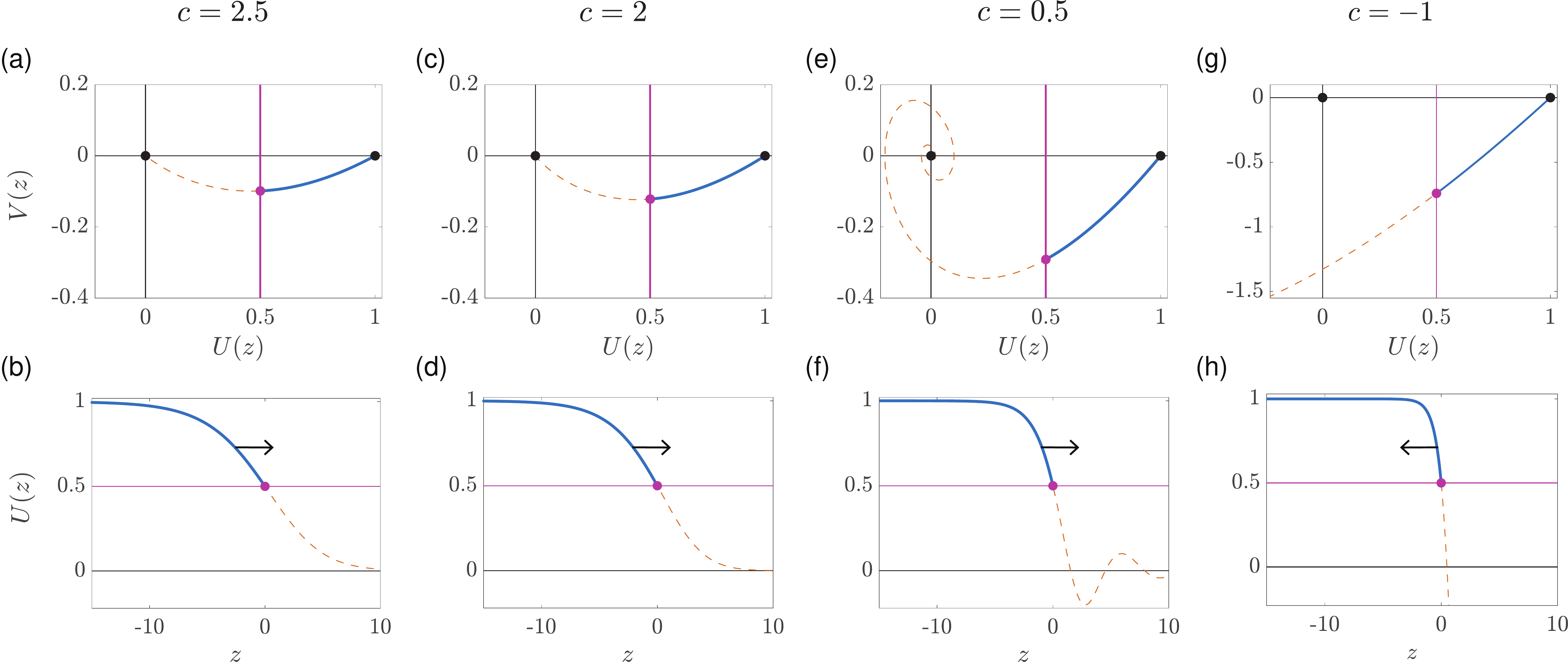}
		\caption{\textbf{Phase planes for invading travelling waves with $u_\mathrm{f}=0.5$.} Phase planes in (a), (c), (e) and (g) show the trajectories corresponding to travelling wave $U(z)$, for $c=2.5, 2, 0.5$ and $-1$ respectively (dashed orange), obtained by solving the dynamical system (\ref{eq:ODEdU})--(\ref{eq:ODEdV}).    Each trajectory is superimposed with a solid blue curve that is obtained from the late-time PDE solutions from Figure \ref{fig:2}. In each phase plane we show the equilibrium points (black disc) and the point at which the trajectory intersects with the vertical line $U=u_\mathrm{f}$ (pink disc).  Results in (b), (d), (f) and (h) show $U(z)$ for each phase plane in (a), (c), (e) and (g) respectively.  These results are shifted so that the moving boundary is at $z=0$. Horizontal lines at $U(z) = 0$ and $U(z) = u_\mathrm{f}$ are superimposed, and the location at which the $U(z)$ curve intersects with $u_\mathrm{f}$ are highlighted (pink disc).}
		\label{fig:3}
	\end{figure}
\end{landscape}

Results in Figure \ref{fig:3}(a), (c), (e) and (g) show the phase plane for $c=2.5, 2, 0.5$ and $-1$, respectively.  In each case the \cbl heteroclinic \cb orbit between $(1,0)$ and $(0,0)$ is shown in dashed pink. In Figure \ref{fig:3}(a) and (b) the heteroclinic orbit enters $(0,0)$ along the dominant eigenvector of the saddle node.  In contrast, in Figure \ref{fig:3}(e) we see the heteroclinic orbit spiraling into $(0,0)$, which is consistent with the linear analysis.   Each phase plane is superimposed with a vertical line at $U(z) = u_\mathrm{f} = 0.5$, and that part of the heteroclinic orbit where $U(z) < u_\mathrm{f}$ is shown as a thick blue line since this is the physically-relevant part of the trajectory corresponds to the travelling wave solution.  In contrast, that part of the trajectory where  $U(z) < u_\mathrm{f}$ is nonphysical, and does not form part of the travelling wave solution~\cite{ElHachem2019}.  Therefore, the travelling wave solutions correspond to a truncated heteroclinic orbit, and this truncation explains why the usual conditions relating to the linearisation about the origin are irrelevant when we consider working in a moving boundary framework.

The role of the Stefan condition in the phase plane is related to the point where the heteroclinic orbit intersects the vertical line where $U(z) = u_\mathrm{f}$. In the phase plane, the Stefan condition corresponds to $c = -\kappa \mathrm{d} U(0)/\mathrm{d} z$, which is equivalent to $c = -\kappa V(0)$.  This means that if the intersection point of the heteroclinic orbit and the vertical line at $u_\mathrm{f}$ is  $(U(0),V(0))$, then  $\kappa = -c/V(0)$, which allows us to calculate $\kappa$ from the phase plane.  For completeness, results in Figure \ref{fig:3}(b), (d), (f) and (h) show $U(z)$ corresponding to the heteroclinic orbits in Figure \ref{fig:3}(a), (c), (e) and (g), respectively.  In these plots we show $U(z)$ superimposed with horizontal lines at $U=0$ (black) and $U=u_\mathrm{f}$ (pink).  The physical part of the travelling wave for $U > u_\mathrm{f}$ and $z < 0$ is shown in solid blue, whereas the nonphysical part of the travelling wave for $z > 0$ is shown in dashed pink.  Indeed, the unphysical part of the $U(z)$ profile in Figure \ref{fig:3}(f) oscillates around $U=0$ as $z \to \infty$.  In all cases we superimpose a pink disc on the point $U=0$ at $z=0$, since this is the point where the Stefan condition applies.

Before proceeding, it is useful to remember the similarities and differences between the time-dependent PDE solutions and the phase plane analysis.  To solve the time-dependent PDE model, Equations (\ref{eq:NondimPDE})--(\ref{eq:ICPDE}), we treat $\kappa$ as an input parameter and the late-time PDE solutions allow us to estimate the wave speed, $c$, which is an output of the model.  In contrast, when we study the heteroclinic orbit in the phase plane, we treat $c$ as an input parameter into (\ref{eq:ODEdU})--(\ref{eq:ODEdV}), and we use the resulting numerical phase plane trajectory to estimate $\kappa=-c/V(0)$, which is an output of the phase plane.  Now that we have demonstrated the relationship between the time-dependent PDE solutions and the phase plane analysis for a range of $c$ and $u_\mathrm{f}$, we will now explore some exact results for special values of $c$ and then develop some insightful perturbation approximations for limiting values of $c$.

\subsection{Stationary wave, $c=0$.} \label{sec:stationarywave}
\begin{figure}[H]
	\centering
	\includegraphics[width=0.85\linewidth]{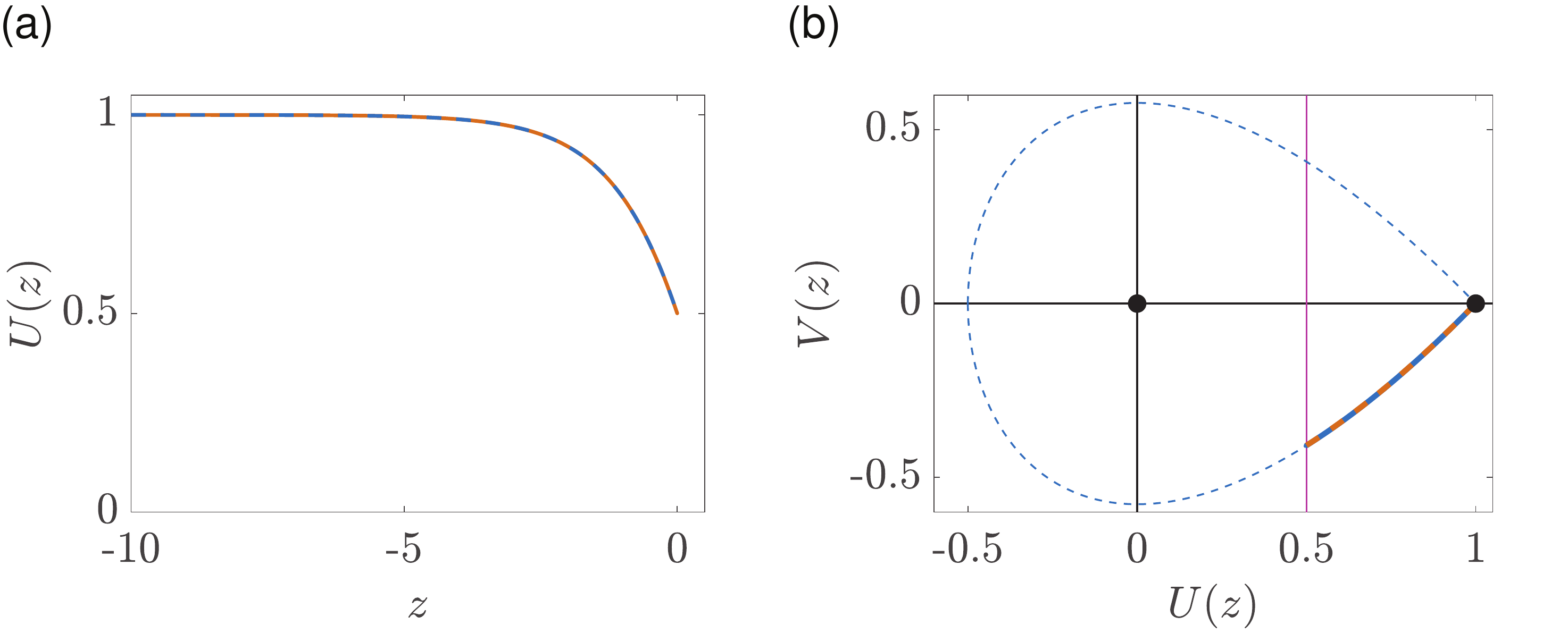}
	\caption{\textbf{Exact solution for $c = 0$ with $u_\mathrm{f} = 0.5$.} (a) Comparison of the exact solution, Equation (\ref{eq:exactUzc0}), (blue) with a late time numerical solution of Equations (\ref{eq:NondimPDE})--(\ref{eq:ICPDE}) (dashed orange) with $\kappa = 0$ and an initial condition with $s(0) = 10$ and $\beta = 1$. (b) Exact phase plane trajectory, Equation (\ref{eq:exactVUc0}) (blue) superimposed with the trajectory obtained by plotting the late-time PDE solution in the phase plane (dashed orange).  The exact homoclinic orbit is given (dashed blue), equilibrium points are highlighted (black discs) along with the vertical line at $U(z)=u_\mathrm{f}$ (pink).}
	\label{fig:4}
\end{figure}

The exact shape of the stationary travelling wave for $c=0$ can be obtained by re-writing Equations (\ref{eq:ODEdU})--(\ref{eq:ODEdV}) as
\begin{equation}\label{eq:odeVU}
\frac{\text{d}V}{\text{d}U} = \frac{-cV - U(1-U)}{V},
\end{equation}
which can be solved when $c=0$, giving
\begin{equation}
V(U) = \pm (1 - U)\sqrt{\dfrac{2U + 1}{3}}. \label{eq:exactVUc0}
\end{equation}
To proceed, we focus on $V(U)<0$. Integrating Equation (\ref{eq:exactVUc0}) with $U(0)=u_\mathrm{f}$ gives an expression for the shape of the stationary wave,
\cbl
\begin{equation}
U(z) =\dfrac{3}{2}\left[\tanh \left(\dfrac{z}{2}-\arctanh\left[\sqrt{\dfrac{2u_\mathrm{f} + 1}{3}}\right]\right)\right]^2-\dfrac{1}{2}, \label{eq:exactUzc0}
\end{equation}
\cb
Results in Figure \ref{fig:4}(a) compare the exact stationary travelling wave solution, Equation (\ref{eq:exactUzc0}), with a late-time numerical solution of Equations (\ref{eq:NondimPDE})--(\ref{eq:ICPDE}) with $\kappa=0$ and $u_{\textrm{f}}=0.5$, showing that the exact result is visually indistinguishable at this scale.  The phase plane for $c=0$ in Figure \ref{fig:4}(b) shows the homoclinic orbit defined by Equation (\ref{eq:exactVUc0}), where for completeness we show both the positive and negative branches.  In this phase plane we show a vertical line at  $u_\mathrm{f}=0.5$, and we also superimpose the late-time numerical solution of Equations (\ref{eq:NondimPDE})--(\ref{eq:ICPDE}) plotted in the phase plane coordinate.  Here we see that the late-time PDE solution is indistinguishable from the truncated homoclinic orbit where $U(z) > u_\mathrm{f}$ and $V(z) < 0$.

\subsection{Solutions with the Painlev\'e property, $c=\pm 5/\sqrt{6}$.} \label{sec:otherexactsol}
\begin{figure}[H]
	\centering
	\includegraphics[width=0.95\linewidth]{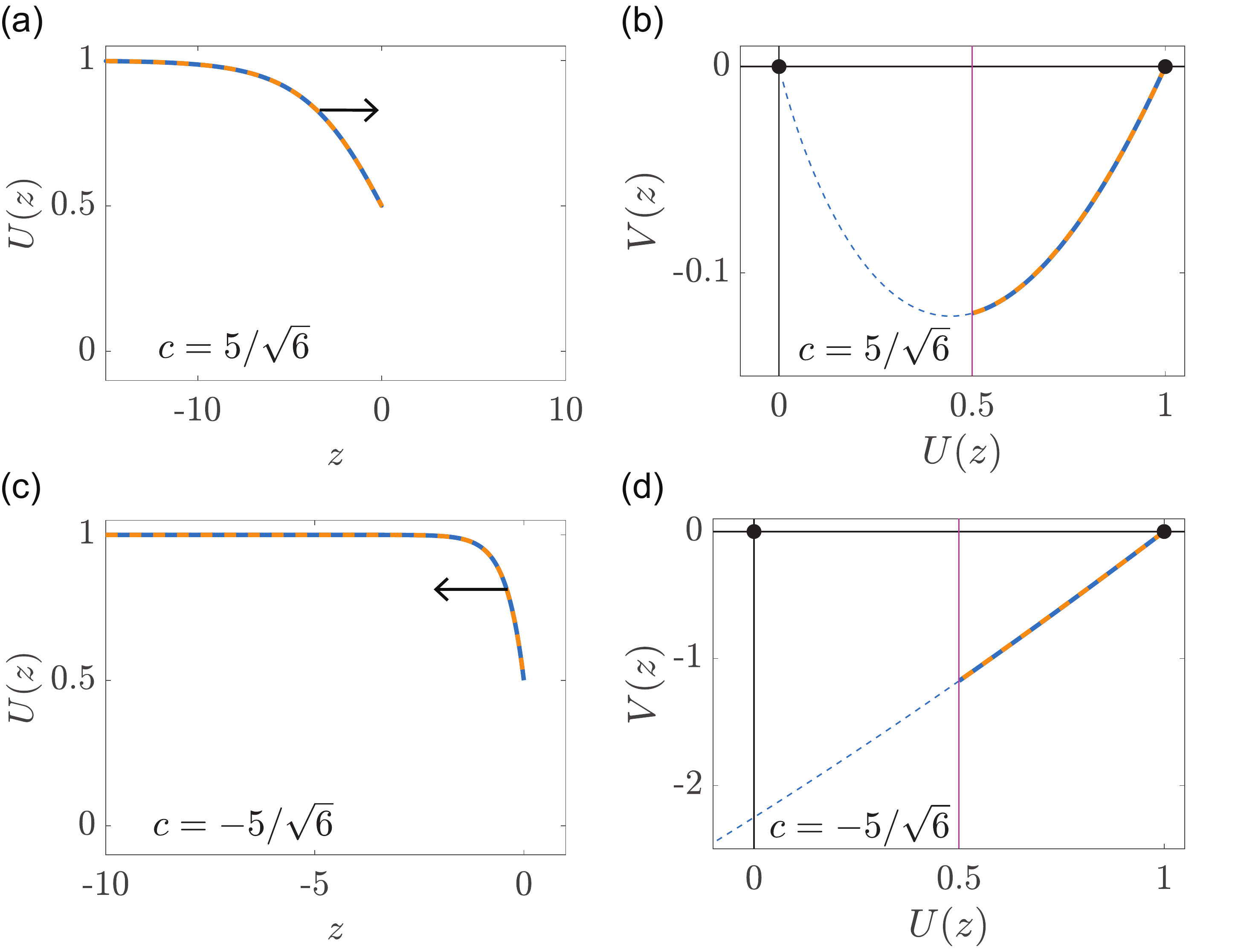}
	\caption{\textbf{Exact solution for $c=\pm 5/\sqrt{6}$ with $u_\mathrm{f} = 0.5$.} (a) and (c) Compare exact solutions given by Equations (\ref{eq:exactUzc5sqrt6}) and (\ref{eq:exactUzminusc5sqrt6}) for  $c=\pm 5/\sqrt{6}$ respectively (blue), with a late time numerical solution of Equations (\ref{eq:NondimPDE})--(\ref{eq:ICPDE}) (dashed orange) with $\kappa = 17.0710$ and $\kappa = -1.7351$, respectively. (b) and (d) Compare the exact trajectories in the phase plane, Equations (\ref{eq:exactVUc5sqrt6})  and (\ref{eq:exactUzminusc5sqrt6})-(\ref{eq:exactVzminusc5sqrt6}) for  $c= \pm 5/\sqrt{6}$, respectively, superimposed with the trajectories obtained by plotting the late-time PDE solution in the phase plane (dashed orange). The phase plane trajectories are given (dashed blue), equilibrium points are highlighted (black discs) along with the vertical line at $U(z)=u_\mathrm{f}$ (pink).}
	\label{fig:5}
\end{figure}

While exact analytic solutions of Equation (\ref{eq:ODEUz}) are unknown for arbitrary values of $c$, it is well known that exact solutions can be written for values of $c$ for which Equation (\ref{eq:ODEUz}) has the  Painlev\'e property,  $c=\pm 5/\sqrt{6}$.  In these cases the solution of Equation (\ref{eq:ODEUz}) can be written in terms of the Weiersta{\ss} p-function~\cite{Ablowitz1979,McCue2021} , and in the case of $c=5/\sqrt{6}$ it is remarkable that this solution can be written very simply in terms of exponential functions~\cite{Kaliappan1984,Murray02},
\begin{equation}
U(z) = \left[1 + \left(-1 + \sqrt{u_{\textrm{f}}} \right)\textrm{e}^{z/\sqrt{6}}\right]^{-2}, \label{eq:exactUzc5sqrt6}
\end{equation}
which corresponds to
\begin{equation}
V(U) = -\dfrac{2U^{3/2}}{\sqrt{6}}\left(\sqrt{\dfrac{1}{U}} - 1\right). \label{eq:exactVUc5sqrt6}
\end{equation}
These two expressions allow us to plot the heteroclinic orbit in the phase plane and to derive an expression for $\kappa = -c/V(u_\mathrm{f})$, giving
\begin{equation}
	\kappa = \dfrac{15}{6u_\mathrm{f}^{3/2}\left(\sqrt{\dfrac{1}{u_\mathrm{f}}}-1\right)}. \label{eq:kappac5sqrt6}
\end{equation}
Results in Figure \ref{fig:5}(a) \cbl show \cb the exact travelling wave solution for $c=5/\sqrt{6}$ and $u_\mathrm{f}=0.5$ superimposed on a late-time PDE solution, showing that the two travelling wave profiles are indistinguishable at this scale.  The corresponding phase plane in Figure \ref{fig:5}(b) compares the exact  heteroclinic orbit with the physically-relevant part of that orbit where $U > u_\mathrm{f}$ from the late-time PDE solution.  The match between the exact result and the numerically-generated phase plane trajectory is excellent.  We note that Equation (\ref{eq:kappac5sqrt6}) allows us to explore how $\kappa$ varies with $u_\mathrm{f}$, for example setting  $u_\mathrm{f}=0.5$ leads to  $\kappa = 5(2+\sqrt{2}) \approx 17.071$.

For $c=-5/\sqrt{6}$ the exact solution can be written in terms of the Weiersta{\ss} p-function~\cite{McCue2021},
\begin{equation}
	U(z) = \mathrm{e}^{2z/\sqrt{6}}\wp\left(\mathrm{e}^{z/\sqrt{6}}-k;0;g_3\right), \label{eq:exactUzminusc5sqrt6}
\end{equation}
giving
\begin{equation}
V(z)=\frac{1}{\sqrt{6}}\mathrm{e}^{2z/\sqrt{6}}\left[2\wp\left(\mathrm{e}^{z/\sqrt{6}}-k;0;g_3\right)+\mathrm{e}^{z/\sqrt{6}}\wp'\left(\mathrm{e}^{z/\sqrt{6}}-k;0;g_3\right)\right], \label{eq:exactVzminusc5sqrt6}
\end{equation}
where the two constants $k$ and $g_3$ are obtained by solving Equation (\ref{eq:exactUzminusc5sqrt6}) with $U(0)=u_\mathrm{f}$ and $-2\pi k g_3^{1/6} = \Gamma(1/3)$~\cite{Ablowitz1979}, where $\Gamma(x)$ is the Gamma function. Results in Figure \ref{fig:5}(c) \cbl show \cb the exact travelling wave solution for $c=-5/\sqrt{6}$ and $u_\mathrm{f}=0.5$ superimposed on a late-time PDE solution, and we see the two profiles are indistinguishable at this scale.  The corresponding phase plane in Figure \ref{fig:5}(d) compares exact phase plane trajectory with the physically-relevant part of the numerically-generated trajectory where $U > u_\mathrm{f}$.  Again the match between the exact result and numerical result is excellent.  As before, the exact solution provides insight into the relationship between $\kappa$ and $u_\mathrm{f}$ by setting $U(\alpha)-u_\mathrm{f}=0$ for $\alpha$ and then calculating $\kappa=-5/[\sqrt{6}V(\alpha)]$.  For example, with $u_\mathrm{f}=0.5$ we have $\kappa =-1.7351$.

\subsection{Slow travelling waves} \label{sec:slowtravwaves}
\begin{figure}[H]
	\centering
	\includegraphics[width=0.95\linewidth]{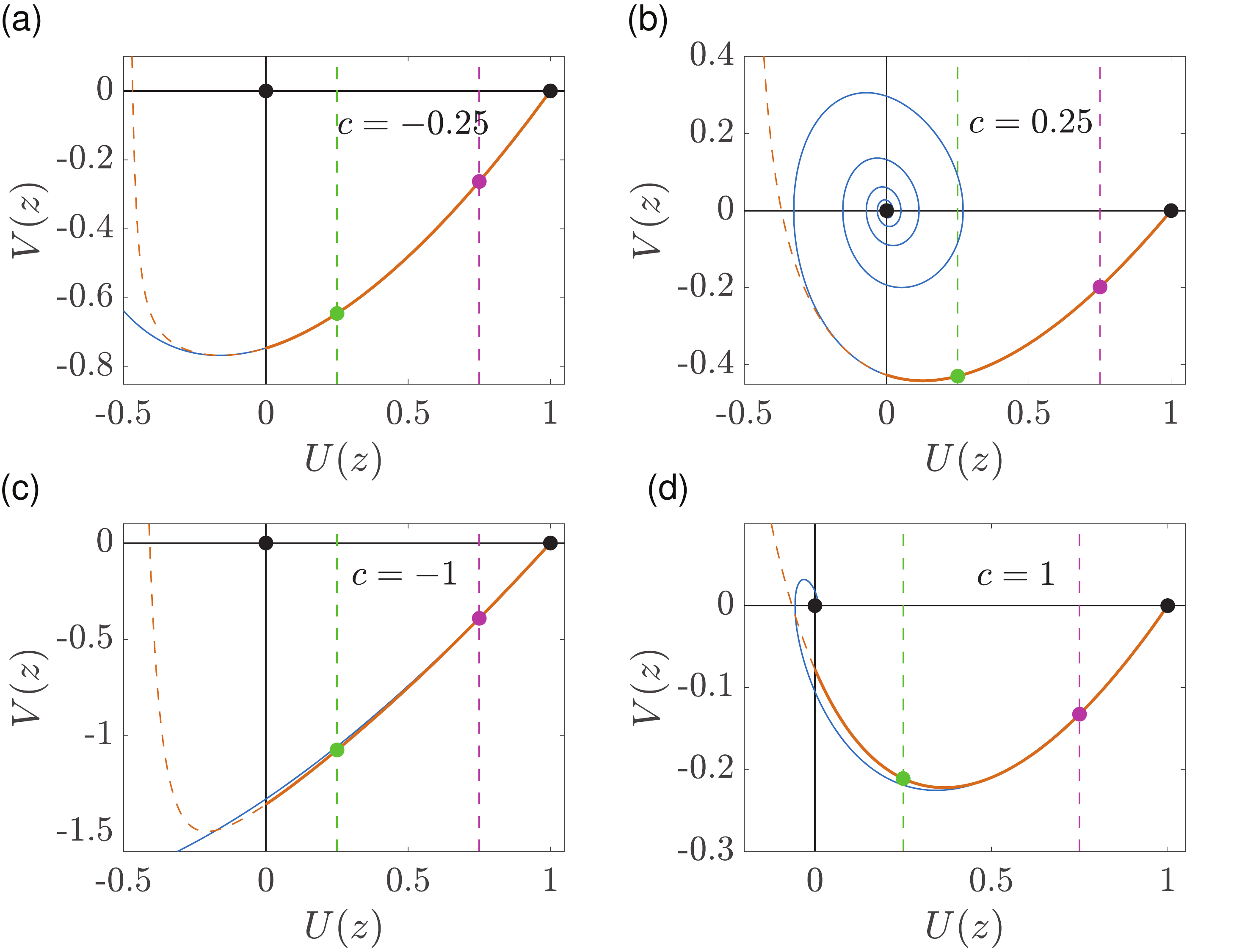}
	\caption{\textbf{Perturbation solutions for $|c| \ll 1$.}  (a)--(d) show phase planes for $c=\mp 0.25$ and $\mp 1.00$, respectively.  Numerical solution of Equations (\ref{eq:ODEdU})--(\ref{eq:ODEdV}) (blue) are superimposed on the perturbation solutions (orange).  The intersection of the perturbation solutions with vertical lines at $U(z)=u_\mathrm{f}=0.25$ and $U(z)=u_\mathrm{f}=0.75$ are highlighted (green and pink discs). Equilibrium points are shown with black discs.}
	\label{fig:6}
\end{figure}

We now build upon the previous results for the stationary wave, $c=0$, to develop insightful approximations for slowly invading or slowly retreating travelling wave solutions.  Seeking a perturbation solution for $|c| \ll 1$, we substitute  $V(U) =  \displaystyle \sum_{n=0}^{\infty} c^n V_n(U)$ into Equation (\ref{eq:odeVU}) to give,
\begin{align}
\label{eq:ODEV0}
&\frac{\text{d}V_0}{\text{d}U}V_0 + U(1-U) = 0,\\
&\frac{\text{d}}{\text{d}U}(V_1 V_0) + V_0 = 0, \label{eq:ODEV1}\\
&\frac{\text{d}}{\text{d}U}(V_2 V_0) + V_1\left(\frac{\text{d}V_1}{\text{d}U} +1\right) = 0, \label{eq:ODEV2}
\end{align}
with boundary conditions $V_0(1) = V_1(1) = V_2(1) = 0$.
The solutions of these differential equations are
\begin{align}
\label{eq:V0csmall}
V_0(U) &= \dfrac{\sqrt{3(2U+1)}(U-1)}{3},\\
V_1(U) &= \dfrac{-(U-2)(2U+1)^{3/2}-3\sqrt{3}}{5(U-1)\sqrt{2U+1}}, \label{eq:V1csmall}\\
\begin{split}
V_2(U) &=  \left(\dfrac{-18\sqrt{3}}{25(2U+1)^{3/2}(U-1)\left(\sqrt{3(2U+1)}-3\right)^2\left(\sqrt{3(2U+1)}+3\right)^2}\right)\\
&\Bigg(\left[\dfrac{(U - 1)^2(2U +1)}{2}\right]\ln\left[\dfrac{(U-1)\left(\sqrt{3(2U+1)}+3\right)}{6\left(\sqrt{3(2U+1)}-3\right)}\right]\\
&-2U^3(6U^2-15U+20)+15U(U+2)+31+6(U-2)(2U+1)\sqrt{3(2U+1)}\Bigg). \label{eq:V2csmall}
\end{split}
\end{align}
We now compare the accuracy of this $\mathcal{O}(c^3)$ perturbation solutions in Figure \ref{fig:6} for $c= \pm 0.25$ and $c = \pm 1$.  The numerical solution of the dynamical system in each phase plane is shown in blue, whereas the perturbation solution is shown in orange.   In all cases we include vertical lines at $u_\mathrm{f}=0.75$ (pink) and $u_\mathrm{f}=0.25$ (green) to illustrate the fact that the accuracy of the perturbation solution depends upon $u_\mathrm{f}$ as well as $c$.  For example, in Figure \ref{fig:6}(d) for $c=1$ we see that the numerically-generated phase plane trajectory and the perturbation solution are visually indistinguishable for $U > 0.75$ meaning that the perturbation solution is very accurate for $u_\mathrm{f}=0.75$.  In contrast, we see some visual discrepancy between the numerically-generated phase plane trajectory and the perturbation solution for smaller values of $U$, which means that the accuracy of the perturbation solution is reduced for $u_\mathrm{f}=0.25$.  Nonetheless, for all values of $c$ in Figure \ref{fig:6} the perturbation solution is very close to the numerically-generated phase plane trajectories. \cbl For completeness we compare $\mathcal{O}(c)$, $\mathcal{O}(c^2)$ and $\mathcal{O}(c^3)$ perturbation solutions for $c= \pm 0.25$ and $c = \pm 1$ in Appendix B.\cb

\begin{figure}[H]
	\centering
	\includegraphics[width=0.75\linewidth]{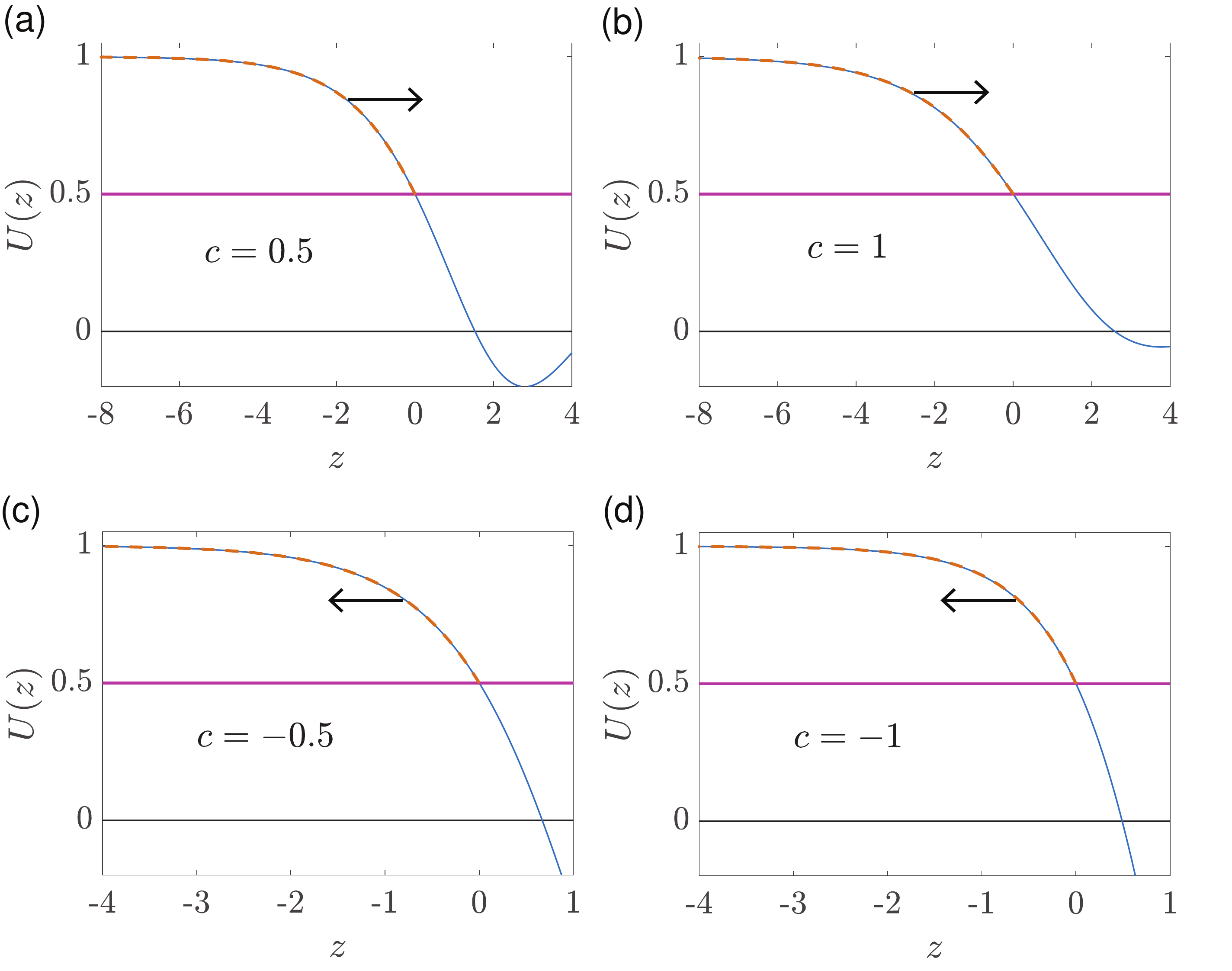}
	\caption{\textbf{Perturbation solutions for slowly invading and retreating travelling waves.}  The shape of travelling wave profile, $U(z)$, obtained using the numerical solution of the phase plane trajectory (blue) is compared with perturbation solution in dashed orange, for $c=0.5$ and $1$ in (a)--(b) and $c=-0.5$ and $-1$ in (c)--(d).}
	\label{fig:7}
\end{figure}

\cbl A comparison of the two solutions in terms of the shape of $U(z)$, where we have numerically integrated the approximate $V(U)$ trajectories in the phase plane, is made in Figure \ref{fig:7}.  Here we compare the numerically-generated phase plane trajectory and the perturbation solution by numerically integrating $V(U)$ using the trapezoid rule.  Plotted in this way, we see that the shape of the travelling wave obtained from the perturbation solution is indistinguishable from the shape of the travelling wave solution generated from the numerically-generated phase plane trajectory for $|c| \le 1$.

\cb

Another way to test the accuracy of the perturbation solution is by comparing our numerical phase plane estimate $\kappa$ with the result obtained from the perturbation solution, $\kappa_{\textrm{p}}$. Evaluating our $\mathcal{O}(c^3)$ perturbation approximation at $U=u_\mathrm{f}$, and then expanding the expression $\kappa = -c/V(u_\mathrm{f})$ in a series gives
\begin{align}
\begin{split}
\kappa_\mathrm{p}&=\dfrac{3}{\sqrt{3(2u_\mathrm{f}+1)}(1-u_\mathrm{f})}c+\dfrac{3}{5}\left[\dfrac{\left(2u_\mathrm{f}^2-3u_\mathrm{f}-2\right)\left(\sqrt{2u_\mathrm{f}+1}\right)+3\sqrt{3}}{(2u_\mathrm{f}+1)^{3/2}(1-u_\mathrm{f})^3}\right] c^2 \\
&-\left(\dfrac{18\sqrt{3}}{25\left(2u_\mathrm{f}+1\right)^{5/2}\left(\sqrt{3}\sqrt{2u_\mathrm{f}+1}+3\right)^2 \left(\sqrt{3}\sqrt{2u_\mathrm{f}+1}-3\right)^2\left(1-u_\mathrm{f}\right)^3}\right)\\
&+\Bigg(90(2u_\mathrm{f}+1)(1-u_\mathrm{f})^2 \ln\left[\dfrac{(u_\mathrm{f}-1)\left(\sqrt{3(2u_\mathrm{f}+1)}+3\right)}{6\left(\sqrt{3(2u_\mathrm{f}+1)}-3\right)}\right]+12u_\mathrm{f}^5-30\left(u_\mathrm{f}^4+6{u_\mathrm{f}}^3\right)\\
&+5\left(39u_\mathrm{f}^2 + 42u_\mathrm{f}\right) +279+54(2u_\mathrm{f}+1)(u_\mathrm{f}-2)\sqrt{3(2u_\mathrm{f}+1)} \Bigg) c^3 + \mathcal{O}\left(c^4\right),
\end{split}
\label{eq:kappacOc2csmall}
\end{align}
\cbl which can be used to estimate $\kappa$ provided we have experimental estimates of $c$ and $u_\mathrm{f}$. \cb

\begin{figure}[H]
	\centering
	\includegraphics[width=1\linewidth]{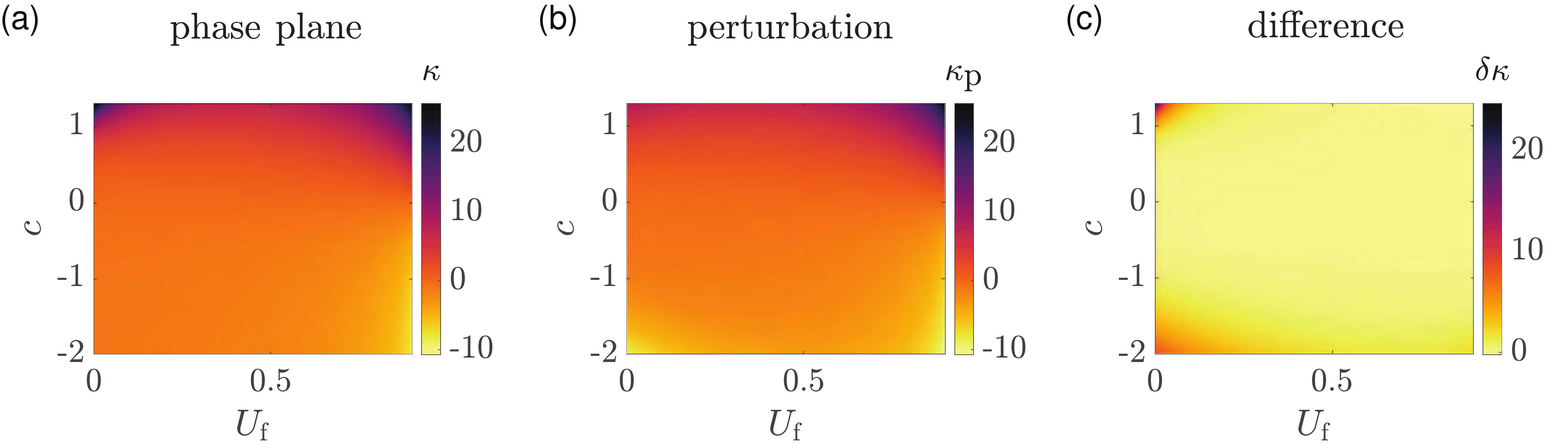}
	\caption{\textbf{$\kappa$ as a function of $c$ and $u_\mathrm{f}$ for $|c| \ll 1$.} (a) Heat map showing $\kappa$ as a function of $c$ and $u_\mathrm{f}$ where estimates of $\kappa$ are obtained by solving Equations (\ref{eq:ODEdU})--(\ref{eq:ODEdV}) in the phase. (b) Heat map showing $\kappa_{\textrm{p}}$ from the perturbation solution, Equation (\ref{eq:kappacOc2csmall}). (c) Difference between the phase plane and perturbation estimates of $\kappa$, $\delta \kappa = \kappa - \kappa_{\textrm{p}}$.}
	\label{fig:8}
\end{figure}

Figure \ref{fig:8}(a) shows a heat map of $\kappa$ as a function of $c$ and $u_\mathrm{f}$ in the interval $-2 < c < 1$ obtained from the phase plane.  The heat map in Figure \ref{fig:8}(b) shows the same result obtained from the perturbation solution, Equation (\ref{eq:kappacOc2csmall}).  The numerically-generated phase plane estimates are difficult to distinguish from the perturbation results, so we plot a heat map of $\delta \kappa = \kappa - \kappa_{\textrm{p}}$ in Figure \ref{fig:8}(c) showing that the difference is small everywhere except for near  $u_\mathrm{f}=0$.

\subsection{Fast retreating travelling waves} \label{sec:fastrecedingtravwaves}
\begin{figure}[H]
	\centering
	\includegraphics[width=1\linewidth]{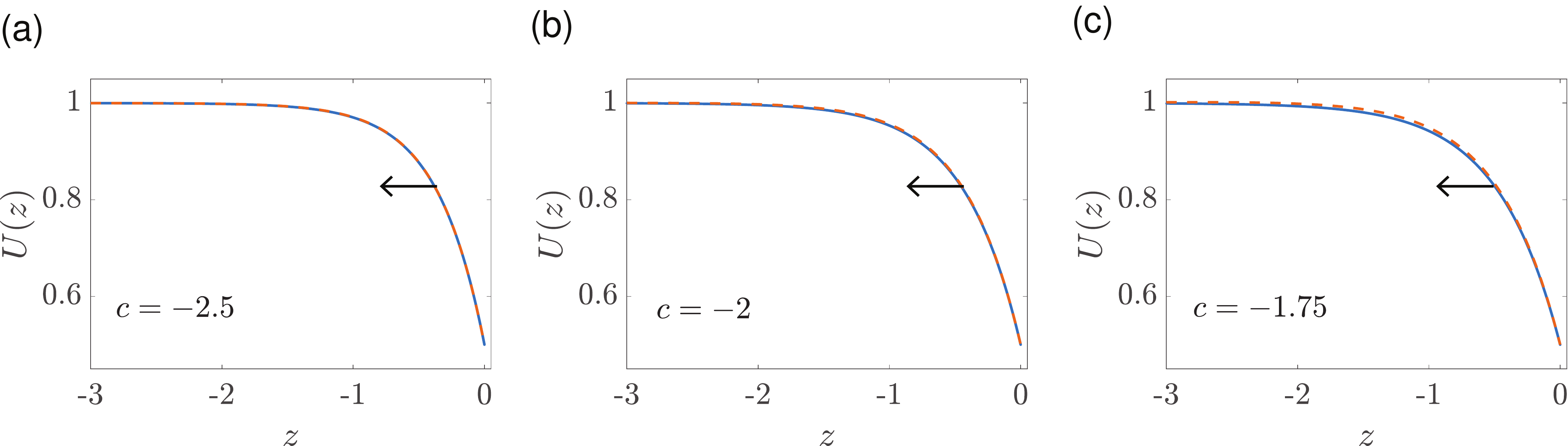}
	\caption{\textbf{Perturbation solution for fast retreating travelling waves, $c \to -\infty$.} (a)--(c) Perturbation solutions showing the shape of travelling waves for $c = -2.5, -2$ and $-1.75$, respectively (dashed orange) superimposed on late-time numerical solutions of Equations (\ref{eq:NondimPDE})--(\ref{eq:ICPDE}) (blue).}
	\label{fig:9}
\end{figure}

We now examine fast retreating travelling wave solutions, \cbl  $|c| \gg 1$\cb, by re-writing the governing boundary value problem as
\begin{equation} \label{eq:ODEUzepsilon}
\dfrac{1}{c}\frac{\mathrm{d}^2 U}{\mathrm{d} z^2} + \frac{\mathrm{d} U}{\mathrm{d} z} + \dfrac{1}{c}U(1-U) = 0,  \quad  -\infty < z < 0,
\end{equation}
which is singular as $c \to -\infty$.  To address this problem we construct a matched asymptotic expansion by treating $1/c$ as a small parameter~\cite{Murray1984}.  The boundary conditions for this problem are \cbl $U(0)=u_\mathrm{f}$ \cb and $U(z) \to 1$ as $z \to -\infty$.  Setting $1/c =0$ and solving the resulting ODE gives the outer solution $U(z) = 1$, which matches the boundary condition as $z \rightarrow -\infty$. To construct the inner solution near $z=0$, we rescale the independent variable $\zeta=zc$. Therefore, in the boundary layer we have
\begin{equation} \label{eq:ODEUxiepsilon}
\frac{\mathrm{d}^2 U}{\mathrm{d} \zeta^2} + \frac{\mathrm{d} U}{\mathrm{d} \zeta} + \dfrac{1}{c^2} U(1-U) = 0,  \quad  0 < \zeta < \infty .
\end{equation}
Substituting $U(\zeta) = \displaystyle  \sum_{n=0}^{\infty} c^{-2n} U_n(\zeta)$ into Equation (\ref{eq:ODEUxiepsilon}) gives
\begin{align} \label{eq:ODEU0}
&\frac{\mathrm{d}^2 U_0}{\mathrm{d} \zeta^2} + \frac{\mathrm{d} U_0}{\mathrm{d} \zeta} = 0,\\
\label{eq:ODEU1}
&\frac{\mathrm{d}^2 U_1}{\mathrm{d} \zeta^2} + \frac{\mathrm{d} U_1}{\mathrm{d} \zeta} + U_0(1-U_0) = 0,\\
\label{eq:ODEU2}
&\frac{\mathrm{d}^2 U_2}{\mathrm{d} \zeta^2} + \frac{\mathrm{d} U_2}{\mathrm{d} \zeta} + U_1(1-2U_0) = 0,\\
\label{eq:ODEU3}
&\frac{\mathrm{d}^2 U_3}{\mathrm{d} \zeta^2} + \frac{\mathrm{d} U_3}{\mathrm{d} \zeta} + U_2(1-2U_0) - U_1^2 = 0,
\end{align}
where $U_0(0) = u_\mathrm{f}$ and $U_{1}(0) = U_{2}(0) = U_{3}(0)= 0$, and $U_0(\zeta) \to 1$, $U_1(\zeta) \to 0$, $U_2(\zeta) \to 0$, and $U_3(\zeta) \to 0$ as  $\zeta \rightarrow \infty$.
The solution of these boundary value problems are
\begin{align}
U_0(\zeta)&= (u_\mathrm{f}-1)\textrm{e}^{-\zeta}+1 \label{eq:perturbationU0zeta}, \\
U_1(\zeta)&=\left(\dfrac{u_\mathrm{f}-1}{2}\right)\left[(u_\mathrm{f}-1)\textrm{e}^{-2\zeta}+(-2\zeta - u_\mathrm{f}+1)\textrm{e}^{-\zeta}\right], \label{eq:perturbationU1zeta}\\
\begin{split}
U_2(\zeta)&=\left(\dfrac{u_\mathrm{f}-1}{12}\right)\left(\left[6\zeta(\zeta+1+u_\mathrm{f})+4u_\mathrm{f}^2+7u_\mathrm{f}-11\right]\textrm{e}^{-\zeta}\right.\\
&\left.+(u_\mathrm{f}-1)\left[-3(4\zeta+2u_\mathrm{f}+3)\textrm{e}^{-2\zeta}+2(u_\mathrm{f}-1)\textrm{e}^{-3\zeta}\right]\right), 		
\end{split}
\label{eq:perturbationU2zeta}\\
\begin{split}
U_3(\zeta)&= \left(\dfrac{u_\mathrm{f}-1}{144}\right)\left[\left(-24\zeta^3-108\zeta^2-12u_\mathrm{f}(3\zeta^2+13\zeta)-4(12u_\mathrm{f}^2+21)\zeta\right.\right.\\
&\left.-37u_\mathrm{f}^3-133u_\mathrm{f}^2-145u_\mathrm{f}+315\right)\textrm{e}^{-\zeta}\\
&+\left(\left[3\left(4\left[12\left(\zeta^2+u_\mathrm{f}\zeta\right)+30\zeta+19(u_\mathrm{f}+1)\right]+22u_\mathrm{f}^2\right)\right]\textrm{e}^{-2\zeta}\right.\\
&+\left.\left.\left[-4(9(2\zeta+u_\mathrm{f})+20)\textrm{e}^{-3\zeta}+7(u_\mathrm{f}-1)\textrm{e}^{-4\zeta}\right](u_\mathrm{f}-1)\right)(u_\mathrm{f}-1)\right].
\end{split}
\label{eq:perturbationU3zeta}
\end{align}
We now compare the accuracy of this $\mathcal{O}(c^{-8})$ perturbation solution in Figure \ref{fig:9} for $c= -2.5, -2$ and $-1.75$ where we superimpose a late-time numerical solution of Equations (\ref{eq:NondimPDE})--(\ref{eq:ICPDE}) onto the perturbation solution in terms of the re-scaled variable, $z=\zeta/c$.  For this comparison we choose $u_\mathrm{f}=0.5$, and we see that the numerical and perturbation solutions are visually indistinguishable for $c = -2.5$.  Results for $c=-2$ and $-1.75$ show some small discrepancy between the numerical and perturbation profiles. \cbl Again, for completeness we compare $\mathcal{O}(c^{-2})$, $\mathcal{O}(c^{-4})$, $\mathcal{O}(c^{-6})$ and $\mathcal{O}(c^{-8})$ perturbation solutions for $c= -2.5, -2$ and $-1.75$ in Appendix B.\cb

Again, we provide a further comparison between the accuracy of the perturbation solution in terms of estimating $\kappa$ from the phase plane and the perturbation solution, which gives
\begin{equation}
\kappa_\mathrm{p} = \dfrac{-\left[1 - \dfrac{u_\mathrm{f}+1}{2c^2} + \dfrac{5u_\mathrm{f}^2+11u_\mathrm{f}+8}{12 c^4}
	+\dfrac{-57u_\mathrm{f}^3-197u_\mathrm{f}^2-281u_\mathrm{f}-185}{144 c^6}\right]}{1-u_\mathrm{f}}  + \mathcal{O}\left(\dfrac{1}{c^8}\right),
\label{eq:kappacOcm6cbig}
\end{equation}
\cbl which again allows us to estimate $\kappa$ provided we have experimental estimates of $c$ and $u_\mathrm{f}$. \cb

\begin{figure}[H]
	\centering
	\includegraphics[width=1\linewidth]{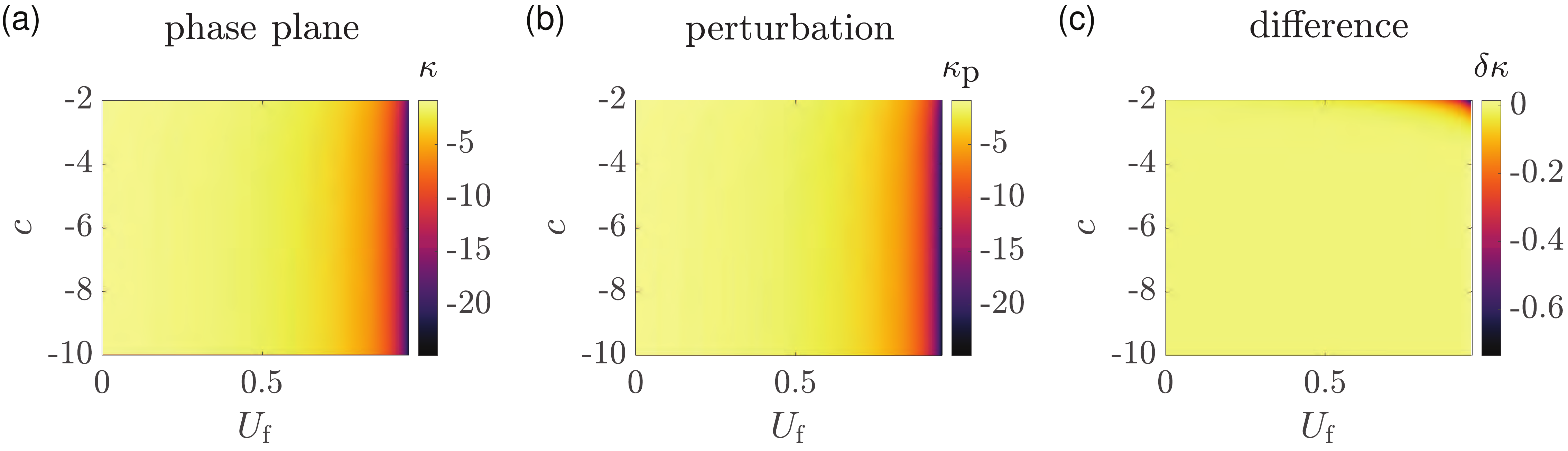}
	\caption{\textbf{$\kappa$ as a function of $c$ and $u_\mathrm{f}$ for $c \to -\infty$.} (a) Heat map showing $\kappa$ as a function of $c$ and $u_\mathrm{f}$ where estimates of $\kappa$ are obtained by solving Equations (\ref{eq:ODEdU})--(\ref{eq:ODEdV}) in the phase plane. (b) Heat map showing $\kappa_{\textrm{p}}$ from the perturbation solution, Equation (\ref{eq:kappacOcm6cbig}). (c) Difference between the phase plane and perturbation estimates of $\kappa$, $\delta \kappa = \kappa - \kappa_{\textrm{p}}$.}
	\label{fig:10}
\end{figure}

Heat maps in Figure \ref{fig:10}(a)--(b) compare numerical estimates of $\kappa$ from the phase plane with the perturbation result, Equation (\ref{eq:kappacOcm6cbig}).  The heat map of $\delta \kappa = \kappa - \kappa_{\textrm{p}}$ in Figure \ref{fig:10}(c) shows that the $\mathcal{O}(c^{-8})$ perturbation solutions leads to extremely accurate solutions for $\kappa$ for $c < -2$ for all $u_\mathrm{f}$.  Equation  (\ref{eq:kappacOcm6cbig}) reveals further information about the existence of travelling wave solutions for this model since we have $\kappa = -1/(1-u_\mathrm{f})$ as $c \to -\infty$.  Indeed, solving Equations (\ref{eq:NondimPDE})--(\ref{eq:ICPDE}) with $\kappa < -1/(1-u_\mathrm{f})$ does not lead to constant speed, constant shape travelling wave solutions.  Instead, for these cases the time-dependent solutions appear to undergo blow-up, as explored in~\cite{McCue2021b}.

\subsection{Fast invading travelling waves} \label{sec:fastinvadingtravwaves}
In Section \ref{sec:fastrecedingtravwaves} we saw that retreating travelling waves become increasingly steep as $c \to -\infty$.  In this section we make use of the fact that, as noted by Murray~\cite{Murray02}, invading travelling waves become increasingly flat as $c \to \infty$.  This means that $V \to 0$ as $c \to \infty$.  Following Canosa~\cite{Canosa1973}, we re-write Equation (\ref{eq:odeVU}) in terms of the re-scaled variable, $\tilde{V} = c V$, giving
\begin{equation}
\dfrac{\tilde{V}}{c^2}\dfrac{\textrm{d} \tilde{V}}{\textrm{d} U} + \tilde{V} + U(1-U).
\end{equation}
Assuming a solution of the form $\tilde{V}(U) = \displaystyle \sum_{n=0}^{\infty} c^{-2n} \tilde{V}_n(U)$, we obtain
\begin{align}
\label{eq:V1canosa}
\tilde{V}_0(U) &= U^2-U, \\
\tilde{V}_1(U) &= - \tilde{V}_0(U)\dfrac{\text{d} \tilde{V}_0(U)}{\text{d} U} =-(2U^3-3U^2+U),\\
\tilde{V}_2(U) &= - \tilde{V}_0(U)\dfrac{\text{d} \tilde{V}_1(U)}{\text{d} U}-\tilde{V}_1(U)\dfrac{\text{d} V_0(U)}{\text{d} U} =-2(5U^4-10U^3+6U^2-U),\label{eq:V2canosa}
\end{align}
which can also be written in terms of the original variable by remembering that $V = \tilde{V}/c$.

\begin{figure}[H]
	\centering
	\includegraphics[width=1\linewidth]{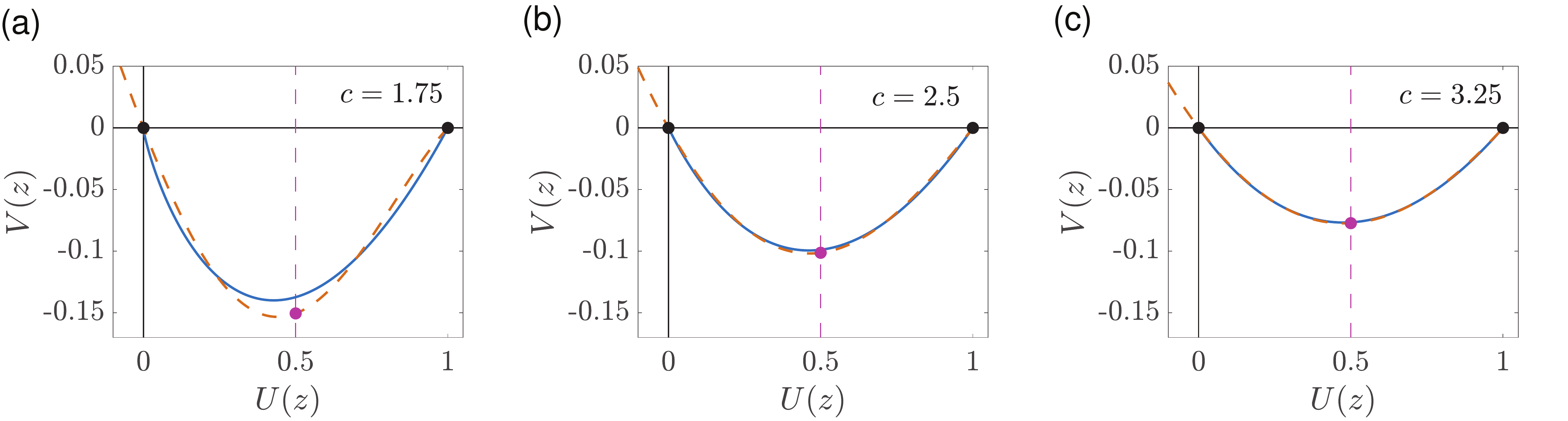}
	\caption{\textbf{Phase plane perturbation solutions for fast retreating travelling waves, $c \to -\infty$.} (a)--(c) Phase plane for $c = 1.75, 2.5$ and $3.25$.  Numerical solutions of  Equations (\ref{eq:ODEdU})--(\ref{eq:ODEdV}) (blue) are superimposed on the perturbation solutions (dashed orange).  The intersection of the perturbation trajectory with the vertical line at $U(z)=u_\mathrm{f}=0.5$ is highlighted (pink disc) and the equilibrium points also highlighted (black discs).}
	\label{fig:11}
\end{figure}

Results in Figure \ref{fig:11} compare numerically-generated phase plane trajectories with the $\mathcal{O}(c^{-6})$ perturbation solution in the phase plane for $c=1.75, 2.5$ and $3.25$. Here we see that the perturbation solution is very accurate for the two faster travelling wave speeds, but we see a visual discrepancy between the numerically-generated phase plane trajectory and the perturbation solution for $c=1.75$.

\begin{figure}[H]
	\centering
	\includegraphics[width=1\linewidth]{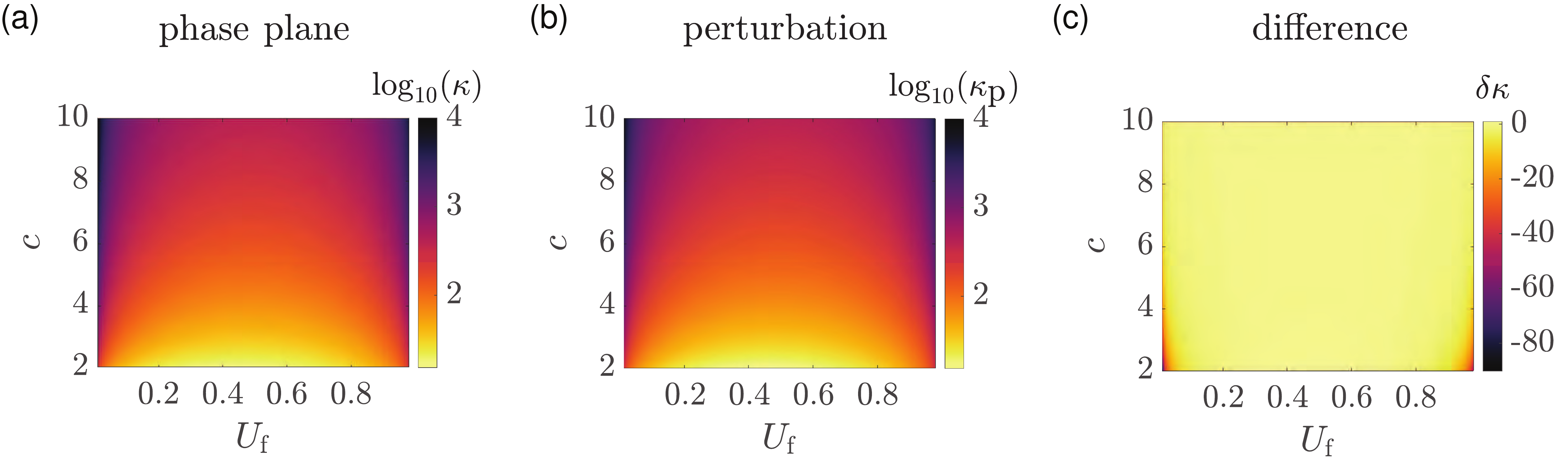}
	\caption{\textbf{$\kappa$ as a function of $c$ and $u_\mathrm{f}$ for $c \to \infty$.} (a) Heat map showing $\kappa$ as a function of $c$ and $u_\mathrm{f}$ where estimates of $\kappa$ are obtained by solving Equations (\ref{eq:ODEdU})--(\ref{eq:ODEdV}) in the phase. (b) Heat map showing $\kappa_{\textrm{p}}$ from the perturbation solution, Equation  (\ref{eq:kappacufcanosa}). (c) Difference between the phase plane and perturbation estimates of $\kappa$, $\delta \kappa = \kappa - \kappa_{\textrm{p}}$.}		
	\label{fig:12}
\end{figure}

As before, another test of the accuracy of the perturbation solution is to compare numerically-generated phase plane estimates of $\kappa$ with the value implied by the perturbation solution, which can be written as
\begin{equation}
\kappa_\mathrm{p} = \dfrac{\left(c^2 + 2u_\mathrm{f}-1 - \dfrac{14u_\mathrm{f}(1-u_\mathrm{f})-3}{c^2} + \dfrac{(2u_\mathrm{f} - 1)[24u_\mathrm{f}(u_\mathrm{f} - 1) + 5]}{c^4}\right)}{u_\mathrm{f}(1-u_\mathrm{f})} + \mathcal{O}\left(\dfrac{1}{c^6}\right).
\label{eq:kappacufcanosa}
\end{equation}
Heat maps in Figure \ref{fig:12}(a)--(b) show $\kappa$ and $\kappa_{\textrm{p}}$ as a function of $c$ and $u_{\textrm{f}}$ using the phase plane and perturbation approaches, respectively.  Visually we see no obvious distinction between the numerical and perturbation approximation of $\kappa$, and this is quantitatively confirmed in Figure \ref{fig:12}(c) where we show a heat map of $\delta \kappa$ which is very close to zero for all $c \ge 2$.

To solve for the shape of the travelling wave as $c \to \infty$ we again follow Canosa~\cite{Canosa1973} and write Equation (\ref{eq:ODEUz}) in terms of the re-scaled coordinate $\xi = z/c$,
\begin{equation}\label{eq:odeUcanosa}
\dfrac{1}{c^2}\dfrac{\text{d}^2U}{\text{d}\xi} + \dfrac{\text{d}U}{\text{d}\xi} + U(1-U) = 0, \quad -\infty < \xi < 0.
\end{equation}
Assuming $U(\xi) = \displaystyle \sum_{n=0}^{\infty} c^{-2n} U_n(\xi)$, and substituting this expansion into Equation (\ref{eq:odeUcanosa}) we obtain
\begin{align}
&\dfrac{\text{d}U_0}{\text{d}\xi} + U_0(1-U_0) = 0,\label{eq:odeU0canosa}\\
&\dfrac{\text{d}U_1}{\text{d}\xi} + \dfrac{\text{d}^2U_0}{\text{d}\xi^2} + U_1(1-2U_0)  =0,\label{eq:odeU1canosa}
\end{align}
with $U_0(0) = u_\mathrm{f}$ and $U_1(0) = 0$, and  $U_0(\xi) = 1$ and $U_1(0)=0$ as $\xi \rightarrow -\infty$.  The solutions of these differential equations are
\begin{align}
U_0(\xi) &= \dfrac{u_\mathrm{f}}{(1-u_\mathrm{f})\textrm{e}^{\xi}+u_\mathrm{f}}, \label{eq:solU0canosa}\\
U_1(\xi) &= \dfrac{u_\mathrm{f}(1 - u_\mathrm{f})\textrm{e}^{\xi}\left(\xi -\ln\left[(1 - u_\mathrm{f})\textrm{e}^{\xi} + u_\mathrm{f}\right]^2\right)}{\left[\left(1 - u_\mathrm{f}\right)\textrm{e}^{\xi} + u_\mathrm{f}\right]^2}. \label{eq:solU1canosa}
\end{align}

\begin{figure}[H]
	\centering
	\includegraphics[width=1\linewidth]{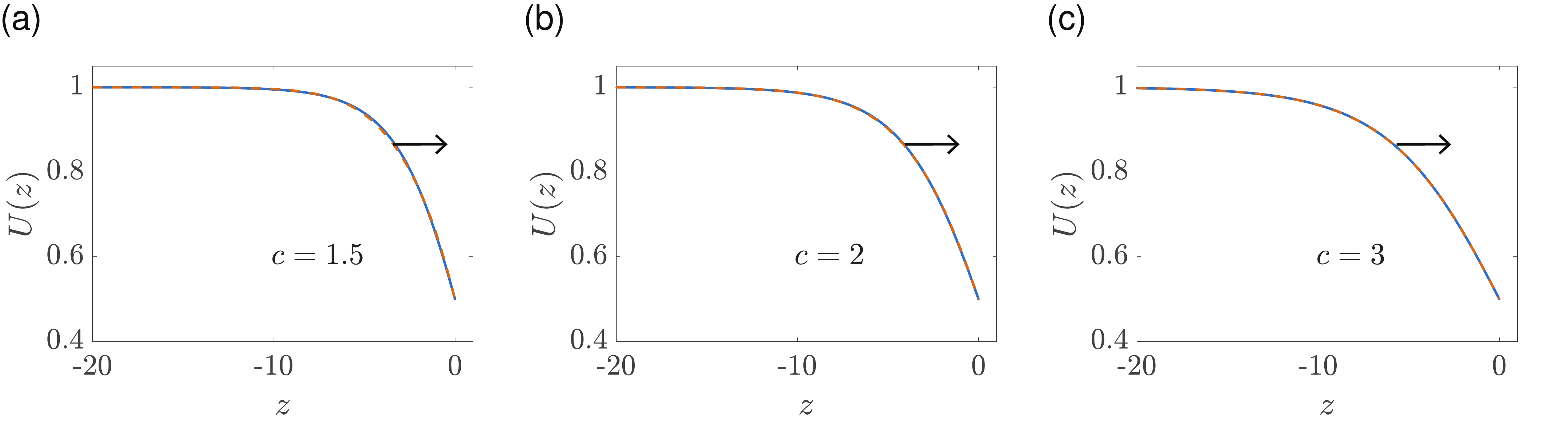}
	\caption{\textbf{Perturbation solution for fast invading travelling waves, $c \to \infty$.} (a)--(c) Perturbation solutions showing the shape of travelling waves for $c = 1.5, 2$ and $3$, respectively (dashed orange) superimposed on late-time numerical solutions of Equations (\ref{eq:NondimPDE})--(\ref{eq:ICPDE}) (blue).}		
	\label{fig:13}
\end{figure}

Results in Figure \ref{fig:13} show late-time numerical solutions of Equations (\ref{eq:NondimPDE})--(\ref{eq:ICPDE}) for $c=1.5, 2$ and $3$, each with $u_{\textrm{f}}=0.5$ in this case.  These numerical travelling wave solutions are superimposed on the $\mathcal{O}(c^{-4})$ perturbation solution derived in this Section and we see that the shape of the travelling waves from perturbation solution provides an excellent approximation of the late-time PDE solutions for all $c$ considered.  This accuracy is remarkable given that the perturbation solutions are valid as $c \to \infty$, yet they match the numerical solutions extremely well for a value as small as  $c=1.5$. \cbl It is interesting to note that all perturbation results here simplify to those given in our previous work which focused on setting $u_{\textrm{f}}=0$~\cite{ElHachem2021a}. \cb

\section{Conclusions and future work}
Despite the widespread popularity of the Fisher-KPP model as a prototype mathematical model of biological invasion, there are some key limitations of travelling wave solutions of this model that are inconsistent with experimental observations of invasive phenomena.  For example, travelling wave solutions of the Fisher-KPP model do not give rise to a well-defined invasion front that arises naturally in many biological scenarios~\cite{Maini2004a,Maini2004b}.  Further,  biologically-relevant initial conditions lead to a very restrictive wave speed.  In this work we show how to reformulate the Fisher-KPP model as a moving boundary problem on $x < s(t)$ with a classical one-phase Stefan condition defining the speed of the moving front.  This approach leads to travelling wave solutions that involve a well-defined sharp front without the complication of introducing a degenerate nonlinear diffusivity.  Furthermore, this moving boundary reformulation of the Fisher-KPP model gives rise to a wide range of travelling wave solutions that move with any speed, $-\infty < c < \infty$.   This is a very interesting result since previous research focusing on retreating travelling wave solutions with $c < 0$ often involves the complication of working with a coupled systems of nonlinear reaction-diffusion equations~\cite{Painter2003,ElHachem2020,Elhachem2021}, whereas here in the moving boundary framework we can simulate retreating travelling wave solutions in a single reaction-diffusion equation.

The important feature in our model \cbl (\ref{eq:NondimPDE})--(\ref{eq:NondimBC}) \cb that leads to a family of travelling wave solutions for all $-\infty < c < \infty$ is the boundary condition $u = u_{\textrm{f}}$ at $x=s(t)$. In previous studies of this Fisher-Stefan model, the parameter $u_{\textrm{f}}$ was fixed to be  $u_{\textrm{f}}=0$, whereas here we focus on $u_{\textrm{f}} \in (0,1)$.  For the case $u_{\textrm{f}}=0$, the wave speeds were restricted to $c < 2$, with the limiting value $c=2$ corresponding to the well-known travelling wave solution to the traditional Fisher-KPP model.  In our model  \cbl (\ref{eq:NondimPDE})--(\ref{eq:NondimBC}) \cb with $u_{\textrm{f}} \in (0,1)$, the speed $c=2$ plays no special role at all.    Another important difference between the cases $u_{\textrm{f}}=0$ and $u_{\textrm{f}} \in (0,1)$ is that for $u_{\textrm{f}}=0$ there is the possibility of population extinction for sufficiently small $s(0)$, leading to the so-called spreading-extinction dichotomy~\cite{Du2010,Simpson2020}.  For  $u_{\textrm{f}} \in (0,1)$, this complication is avoided.

A key limitation of reformulating the Fisher-KPP model as a moving boundary problem  \cbl (\ref{eq:NondimPDE})--(\ref{eq:NondimBC}) \cb is the interpretation and estimation of $\kappa$, which is a leakage parameter that describes how the population is lost $(\kappa > 0)$ or gained $(\kappa < 0)$ at the moving boundary.  Here we seek to address this issue by using a range of exact and approximate perturbation solutions to estimate $\kappa$ as a function of $c$, which is useful because the travelling wave speed is relatively straightforward to measure~\cite{Maini2004a,Maini2004b}.  Our analysis gives three exact values for $\kappa$ when $c=\pm 5\sqrt{6}$ and $c=0$, and our perturbation solutions give expressions for $\kappa$ in various limits.  Comparing our perturbation approximations with numerical estimates from the phase plane, our approximations for $\kappa$ are accurate across the entire range of potential travelling wave speeds, $-\infty < c < \infty$. \cbl While we have compared our perturbation solutions with numerical results, it is possible that further work could be completed to improve the accuracy of these solutions by, for example, introducing a Pad\'{e} approximant~\cite{vanDyke1975}. \cb

While our analysis here focuses on invasion phenomena in one-dimensional geometries where we can obtain several exact and approximate perturbation solutions, future work could involve examining numerical solutions in two-dimensional geometries~\cite{King1999,McCue2003,McCue2005} since this would provide a more realistic description of populations of cells that invade outward from an initially-confined region~\cite{Treloar2014} as well as hole-closing problems that describe the closure of an initial gap in an otherwise uniform population~\cite{McCue2019}.

\paragraph{Data Accessibility} This article has no additional data. Key algorithms used to generate results are available at  \href{https://github.com/ProfMJSimpson/TravellingWaves_ElHachem2021}{GitHub}.
\paragraph{Authors Contributions} All authors conceived and designed the study; M.El-H. performed numerical and symbolic calculations and drafted the article.  All authors edited the article and gave final approval for publication.
\paragraph{Competing Interests} We have no competing interests.
\paragraph{Funding} This work is supported by the Australian Research Council (DP200100177).
%\cbl \paragraph{Acknowledgements} We thank two referees for helpful suggestions. \cb

\newpage
\appendix

\section{Numerical methods}
\label{sec:numericalmethods}

\subsection{Partial differential equations}
To obtain numerical solutions of the Fisher--Stefan equation (\ref{eq:NondimPDE}), we use a boundary fixing transformation $\xi = x / s(t)$ so that we have
\begin{align}\label{eq:FisherKPPmovboundxi}
\frac{\partial u}{\partial t} = \frac{1}{s^2(t)} \frac{\partial^2 u}{\partial \xi^2}+\frac{\xi}{s(t)} \frac{\text{d}s(t)}{\text{d} t} \frac{\partial u}{\partial \xi} + u(1-u),
\end{align}
on the fixed domain, $0 < \xi < 1$.  Here $s(t)$ is the time--dependent length of the domain, and we will explain how we solve for this quantity later. To close the problem we also transform the boundary conditions giving
\begin{align}
&\dfrac{\partial u}{\partial \xi} = 0 \quad \textrm{at} \quad \xi=0,  \label{eq:FS_BC1} \\
&u = u\mathrm{_f} \quad \textrm{at} \quad \xi=1. \label{eq:FS_BC2}
\end{align}

The key to obtaining accurate numerical solutions of equation (\ref{eq:NondimPDE}) is to take advantage of the fact that for many problems we consider $u(x,t)$ varies rapidly near $x=s(t)$, whereas $u(x,t)$ is approximately constant near $x=0$.  Motivated by this we discretize equation (\ref{eq:FisherKPPmovboundxi}) using a variable mesh where the mesh spacing varies geometrically from $\delta \xi_{\textrm{min}} = \xi_{N} - \xi_{N-1} = 1 - \xi_{N-1}$ at $\xi = 1$, to $\delta \xi_{\textrm{max}} = \xi_2 - \xi_1 = \xi_2 - 0$ at $\xi = 0$.  All results in this work are computed with $N=5001$ mesh points with $\delta \xi_{\textrm{min}} = 1 \times 10^{-6}$.  With these constraints we solve for the geometric expansion factor 1.01 using MATLABs \textit{fsolve} function which gives $\delta \xi_{\textrm{max}} = 1.457\times 10^{-3}$.

We spatially discretise equation (\ref{eq:FisherKPPmovboundxi}) on the non-uniform mesh.  At the $i$th internal mesh point we define $h_i^+ = \xi_{i+1} - \xi_{i}$ and $h_i^- = \xi_{i} - \xi_{i-1}$.   For convenience we define $\alpha_i = 1/(h^-[h^+ + h^-])$, $\gamma_i = -1/(h^- h^+)$ and $\delta_i = 1/(h^+[h^+ + h^-])$, which gives
\begin{align}\label{eq:FDDinternalxi}
\dfrac{u_{i}^{j+1} - u_{i}^{j} }{\Delta t} &=  \dfrac{2}{(s^{j})^2} \left[ \alpha_i u_{i-1}^{j+1} + \gamma_i u_{i}^{j+1} + \delta_i u_{i+1}^{j+1}  \right]\notag \\
&+\dfrac{\xi_i}{s^{j}} \left(\dfrac{s^{j+1} - s^{j} }{\Delta t}\right) \left[ - \alpha_i h^+ u_{i-1}^{j+1} + \gamma_i (h^--h^+) u_{i}^{j+1} +\delta_i h^- u_{i+1}^{j+1} \right] \notag \\
& +  u_{i}^{j+1}(1 - u_{i}^{j+1}) ,
\end{align}
for $i = 2, \ldots, N-1$, where $N$ is the total number of spatial nodes in the mesh, and index $j$ represents the time index so that $u_{i}^{j} \approx u(\xi_i, j\Delta t)$.

Discretising the boundary conditions (\ref{eq:FS_BC1})--(\ref{eq:FS_BC2}) gives
\begin{align}
&u_{2}^{j+1}-u_{1}^{j+1} = 0,  \label{eq:FS_BC1a} \\
&u_{N}^{j+1} = u\mathrm{_f}. \label{eq:FS_BC2a}
\end{align}

To advance the discrete system from time $t$ to $t + \Delta t$ we solve the system (\ref{eq:FDDinternalxi})-(\ref{eq:FS_BC2a}), using Newton-Raphson iteration. During each iteration we estimate the position of the moving boundary using the discretised Stefan condition.  Here we define $h_N^+ = \xi_{N} - \xi_{N-1}$, $h_N^- = \xi_{N-1} - \xi_{N-2}$, $\alpha_i = 1/(h^-[h^+ + h^-])$, $\gamma_i = -1/(h^- h^+)$ and $\delta_i = 1/(h^+[h^+ + h^-])$, which gives
\begin{equation}
s^{j+1} = s^{j} - \dfrac{\Delta t  \kappa}{s^j} \left[-\alpha_i h^+ u_{N-2}^{j+1} + \gamma_i (h^- -h^+) u_{N-1}^{j+1} + \delta_i h^- u_\textrm{f} \right].
\label{eq:Lupdatediscretise}
\end{equation}
Within each time step Newton-Raphson iterations continue until the maximum change in the dependent variables is less than the tolerance  $\epsilon$.  All results in this work are obtained by setting $\epsilon = 1 \times 10^{-10}$, and $\Delta t = 1 \times 10^{-3}$, and we find that these values are sufficient to produce grid--independent results.  MATLAB software is available on \href{https://github.com/ProfMJSimpson/TravellingWaves_ElHachem2021}{GitHub} so that these algorithms can be implemented to explore different choices of $\delta \xi_{\textrm{min}}$, $\delta \xi_{\textrm{max}}$, $N$, $\delta t$ and $\epsilon$.  For certain problems in this work we the time--dependent solutions to provide an estimate of the velocity of the moving front, $v$. The estimated velocity is computed as $v = (s^{j+1} - s^{j})/\Delta t$, and we find that $v$ approaches as constant travelling wave speed, $c$,  as $t$ becomes sufficiently large.

\subsection{Phase plane}

To construct the phase planes we solve equations (\ref{eq:ODEdU})-(\ref{eq:ODEdV}) numerically using Heun's method with a constant step size $\textrm{d}z$. In most cases we are interested in examining trajectories that either leave the saddle $(1,0)$ along the unstable manifold. We chose the initial condition on the unstable manifold sufficiently close to $(1,0)$. To choose this point we use the MATLAB \textit{eig} function to calculate the eigenvalues and eigenvectors for the particular choice of c of interest.

\newpage
\section{Additional results} \label{sec:suppresults}

\begin{figure}[H]
	\centering
	\includegraphics[width=0.75\linewidth]{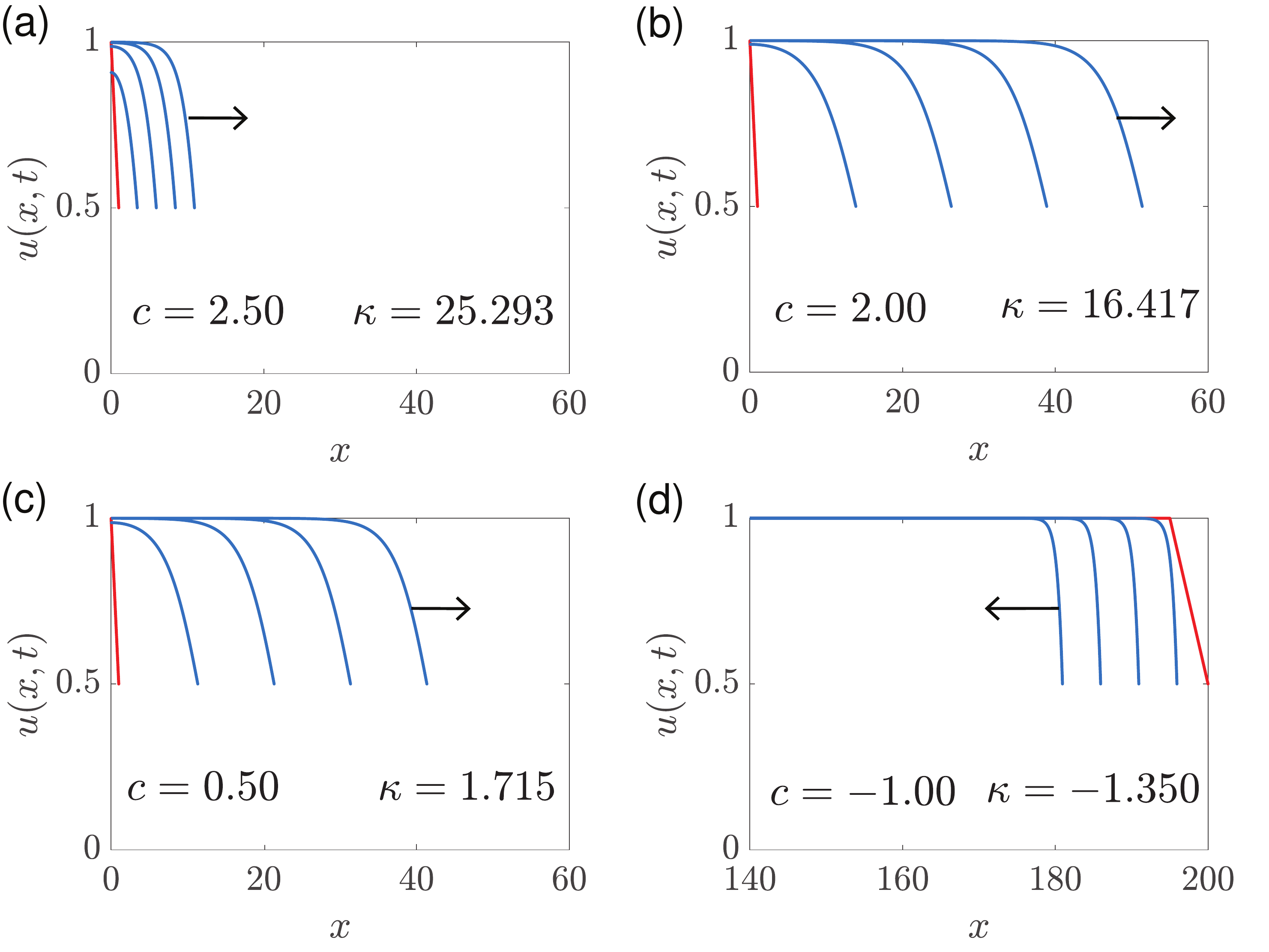}
	\caption{\textbf{Time-dependant solutions of Equations (1.1)--(1.3) for  $u_{\mathrm{f}}=0.5$.} Density profiles $u(x,t)$ are illustrated in blue at times $t=5, 10, 15, 20$. The initial condition is illustrated in red,  where $s(0)=1$ and $\beta = 0$ in (a)--(c), and $s(0)=200$ and $\beta = 195$ in (d). Positive wave speeds $c=0.50,2.00$ and $2.50$ are obtained by $\kappa = 1.715, 16.417$ and $25.293$ and Negative wave speed $c=-1.00$ is obtained by $\kappa = -1.350$. }
	\label{fig:14}
\end{figure}

Additional time-dependent solutions of the moving boundary problem are given in Figure \ref{fig:14} where $u_{\mathrm{f}}=0.5$.  In the main document we show results in Figure \ref{fig:2} for $u_{\mathrm{f}}=0.25$ and $u_{\mathrm{f}}=0.75$, and here we show results for another choice of $u_{\mathrm{f}}$ for completeness.

\begin{figure}[H]
	\centering
	\includegraphics[width=1\linewidth]{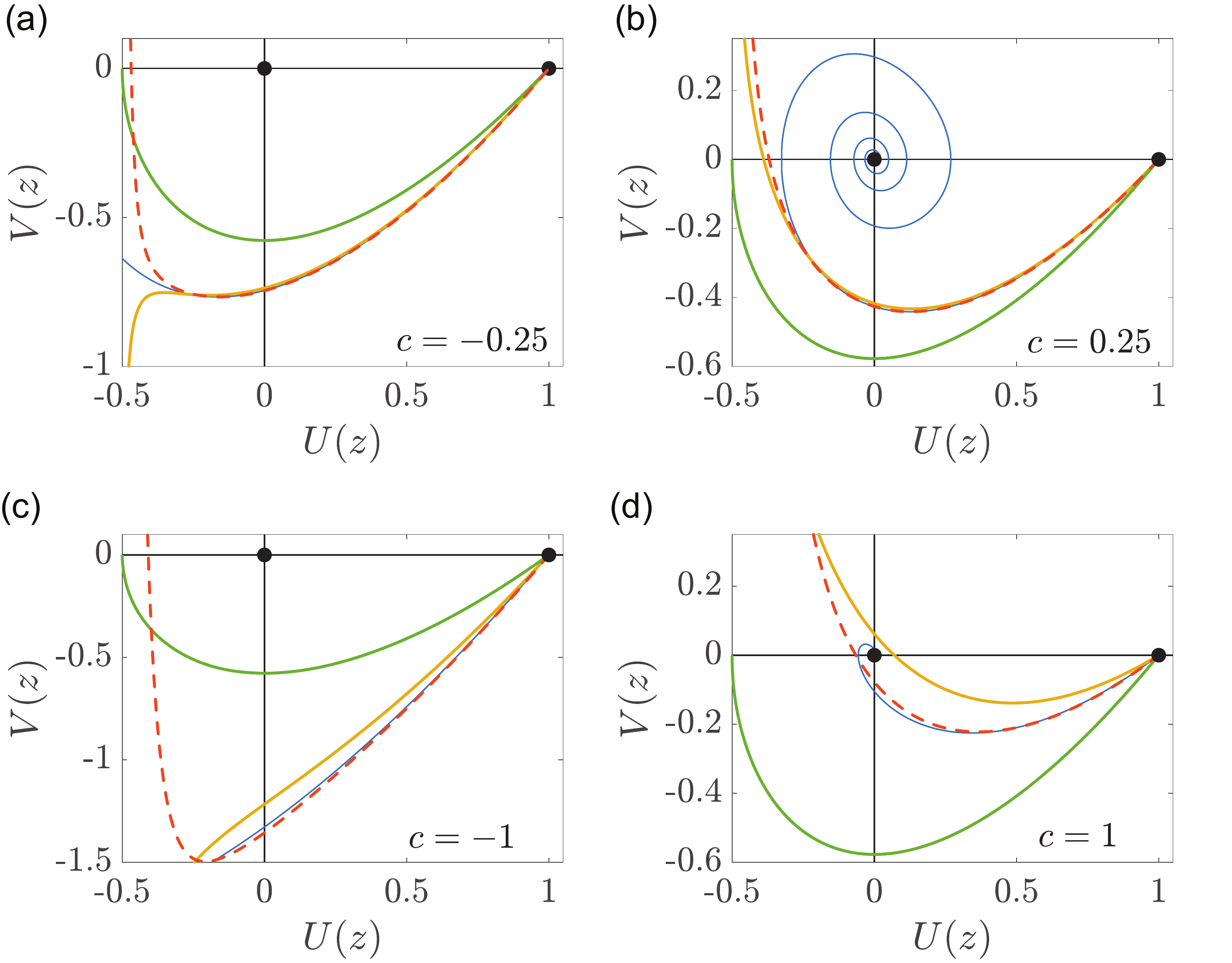}
	\caption{\cbl\textbf{Additional perturbation solutions for $|c| \ll 1$.}  (a)--(d) show phase planes for $c=\mp 0.25$ and $\mp 1.00$, respectively.  Numerical solution of Equations (\ref{eq:ODEdU})--(\ref{eq:ODEdV}) (blue) are superimposed on various perturbation solutions: $\mathcal{O}(c)$ in solid green; $\mathcal{O}(c^2)$ in solid yellow; and,  $\mathcal{O}(c^3)$ in dashed orange. Equilibrium points are shown with black discs.\cb}
	\label{fig:15}
\end{figure}

\cbl Results in Sections \ref{sec:slowtravwaves}--\ref{sec:fastrecedingtravwaves} compare several numerical trajectories in the phase plane with our various perturbation solutions.  These comparisons do not explore the effect of truncation of the perturbation solutions since we always worked with the most terms possible.  Additional results in Figure \ref{fig:15} replicate those in Figure \ref{fig:6} except here we show various perturbation solutions of different order: $\mathcal{O}(c)$ in solid green; $\mathcal{O}(c^2)$ in solid yellow; and, $\mathcal{O}(c^3)$ in dashed orange.  For these particular choices of $c$ we observe the importance of taking higher order terms in the perturbation solutions since the $\mathcal{O}(c)$ perturbation solution is relatively inaccurate in all cases considered.\cb

\begin{figure}[H]
	\centering
	\includegraphics[width=.6\linewidth]{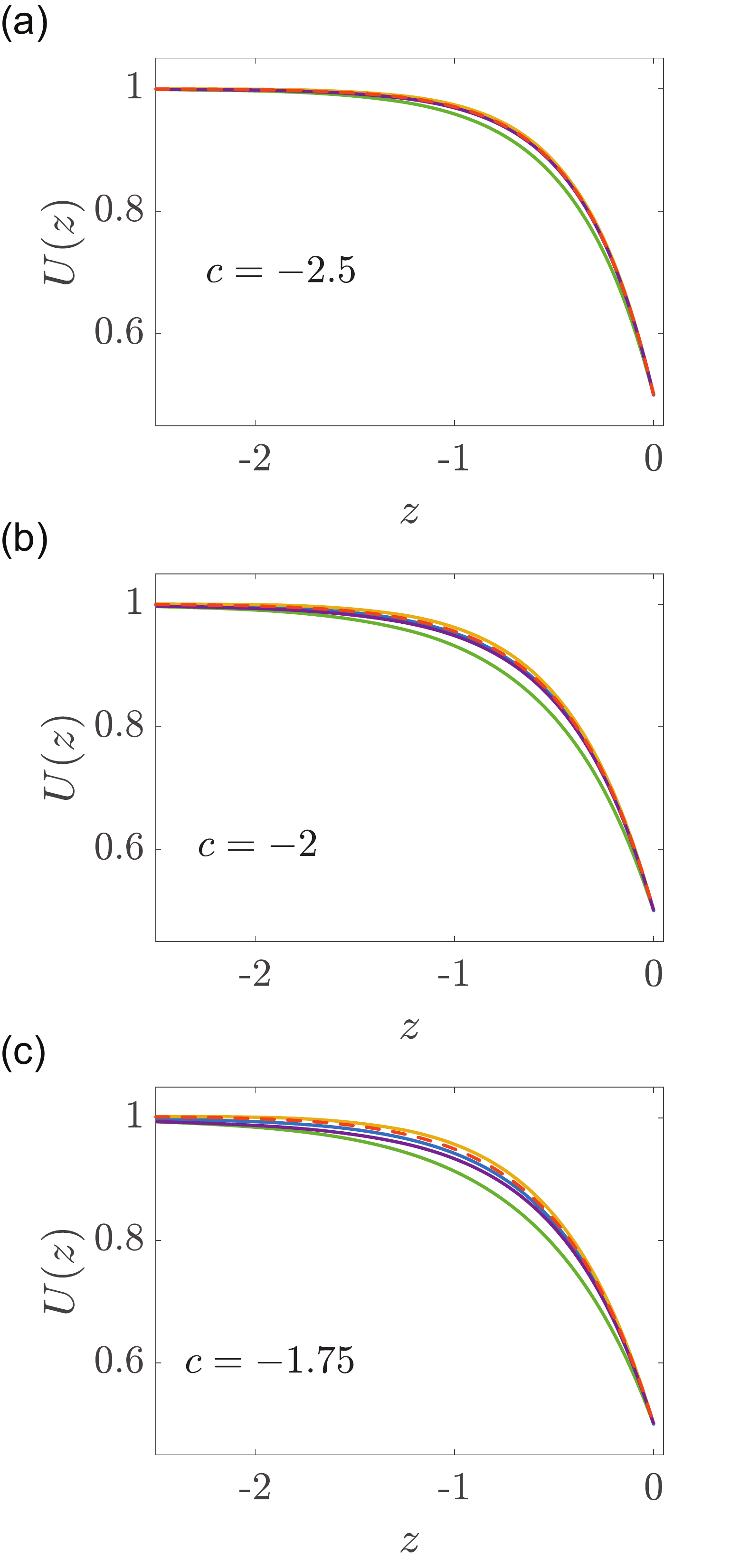}
	\caption{\cbl \textbf{Additional perturbation solution for fast retreating travelling waves, $c \to -\infty$.} (a)--(c) Perturbation solutions of different order of accuracy superimposed on late-time numerical solutions of Equations (\ref{eq:NondimPDE})--(\ref{eq:ICPDE}) in solid blue. Perturbation solutions include: $\mathcal{O}(c^{-2})$ in solid green; $\mathcal{O}(c^{-4})$ in solid yellow; $\mathcal{O}(c^{-6})$ in solid purple; and, $\mathcal{O}(c^{-8})$ in dashed orange.  Results are compared for $c = -2.5, -2$ and $-1.75$, respectively, as indicated. \cb}
	\label{fig:16}
\end{figure}

\cbl Results in Figure \ref{fig:16} replicate those in Figure \ref{fig:11} except here we show various perturbation solutions of different order: $\mathcal{O}(c^{-2})$ in solid green; $\mathcal{O}(c^{-4})$ in solid yellow; $\mathcal{O}(c^{-6})$ in solid purple;  and, $\mathcal{O}(c^{-8})$ in dashed orange.  Just like the comparisons in Figure \ref{fig:15}, for these choices of $c$ here we observe the importance of taking higher order terms in the perturbation solutions since the $\mathcal{O}(c^{-2})$ perturbation solution is relatively inaccurate, particularly for $c=-1.75$.\cb

\cb
%%%%%%%%%% Insert bibliography here %%%%%%%%%%%%%%

\end{document}